\documentclass[useAMS,usenatbib]{mn2e}

\usepackage[latin1]{inputenc}
\usepackage[T1]{fontenc}
\usepackage[dvips]{graphicx}

\usepackage{graphicx}
\usepackage{amssymb}
\usepackage{subfigure}

\title[ON THE STABILITY OF THE SATELLITES OF ASTEROID (87) SYLVIA]{ON THE STABILITY OF THE SATELLITES OF ASTEROID 87 SYLVIA}
\author[O. C. Winter et. al] {O. C.
Winter$^{1}$\thanks{ocwinter@pq.cnpq.br}, L.A.G. Boldrin$^{1}$, E. Vieira Neto$^{1}$, R. Vieira Martins$^{2}$,\\  S.M. Giuliatti Winter$^{1}$, R. S. Gomes$^{2}$, F. Marchis$^{3}$ and P. Descamps$^{4}$ \\
$^{1}$S\~ ao Paulo State University - UNESP, Grupo de Din\^{a}mica Orbital \& Planetologia, Guaratinguet\'{a}, CEP 12.516-410, Brazil.\\
$^{2}$Observat\' orio Nacional, Rua General Jos\' e Cristino, 77, CEP 20921-400, Rio de Janeiro, Brazil\\
$^{3}$ University of California at Berkeley, Department of Astronomy, 601 Campbell Hall, Berkeley, CA 94720, EUA\\
$^{4}$ Institut de M\' ecanique C\' eleste et de Calcul des \' Eph\' em\' erides,  Observatoire de Paris, 75014 Paris, France}

\begin{document}
\maketitle

\label{firstpage}

\begin{abstract}

The triple asteroidal system (87) Sylvia is composed of a 280-km primary 
and two small moonlets named Romulus and Remus (Marchis et al 2005).
Sylvia is located in the main asteroid belt, with semi-major axis of about 
3.49 AU, eccentricity of 0.08 and $11^\circ$ of orbital inclination 
The satellites are in nearly equatorial circular orbits around the primary, with orbital
radius of about 1,360 km (Romulus)  and 710 km (Remus).
In the present work we study the stability of the satellites Romulus and Remus.
In order to identify the effects and the contribution of each perturber
we performed numerical simulations considering a set of different systems.
The results from the 3-body problem, Sylvia-Romulus-Remus, show no significant
variation of their orbital elements. 
However, the inclinations of the satellites present a long period evolution
with amplitude of about $20^\circ$ when the Sun is included in the system.
Such amplitude is amplified to more than $50^\circ$ when Jupiter is included.
 These evolutions are very similar for both satellites.
An analysis of these results show that Romulus and Remus are librating in 
a secular resonance and their longitude of the nodes are locked to each other.
Further simulations show that the amplitude of oscillation of the satellites' 
inclination can reach
higher values depending on the initial values of their longitude of pericentre.
In those cases the satellites get caught in an evection resonance with Jupiter,
their eccentricities grow and they eventually collide with Sylvia.
However, the orbital evolutions of the satellites became completely stable when 
the oblateness of Sylvia is included in the simulations. 
The value of Sylvia's  $J_2$ is about 0.17, which is
 very high. However, even just 0.1\% of this value is enough to keep
the satellite's orbital elements with no significant variation.
\end{abstract}


\section{Introduction}

The first triple asteroidal system discovered was (87) Sylvia (Marchis et al. 2005).
The primary body is about 280 km, while the two small satellites,
Romulus and Remus, are about $18\pm 4$ km and $7\pm 2$ km, respectively.
For simplicity, in this paper we will call the primary of the system (87) Sylvia as ``Sylvia''.
The satellites are in nearly equatorial circular orbits around the primary, 
with orbital radius of about 1,360 km (Romulus)  and 710 km (Remus).
In this paper we study the stability of 
the satellites Romulus and Remus. 
Sylvia is located in the main asteroid belt, with semi-major axis of about 
3.49 AU, eccentricity of 0.08 and $11^\circ$ of orbital inclination.
Therefore, the two main perturbers of the system are the Sun and Jupiter.

In order to identify the effects and the 
 contributions of each relevant perturber of the system we explored 
several different dynamical systems.
We studied the interactions between the two satellites, the perturbation
due to the Sun and due to the planet Jupiter. We identify two secular
resonances that play important roles on the dynamics of such systems.
Finally, we investigate the contribution due to the oblateness
of Sylvia.

This paper has the following structure. 
In the next section we present our numerical simulations considering different
dynamical systems, the 3-, 4- and 5-body problems.
The effects due to the interaction between Romulus and Remus are analysed in Section 3. 
The occurrence of the evection resonance between the satellites and Jupiter and its 
consequences are shown in section 4. 
Section 5 is devoted to the effect of the oblateness
 and finally, in Section 6, we present our conclusions.

\section{Simulations of the 3, 4 and 5-body Problems}

We performed numerical simulations of several different dynamical systems.
In all cases we used the Gauss-radau integrator (Everhart 1985) and considered a 
time span of $5\times 10^4$ years, that corresponds to about $5\times 10^6$ orbital periods of Romulus.
Such time span was found to be suitable in order to capture all the main long term effects
associated to the secular perturbations due to Jupiter.
In order to check the accuracy of the numerical integrations we monitored the total energy of the system.
The relative error was always lower than $10^{-11}$.
The physical and orbital data of the system (87) Sylvia adopted in most of the 
simulations is given in Table 1.

First of all, we looked at the orbital evolution  of Sylvia.
Sylvia's orbit around the Sun, under Jupiter's perturbation, 
presents a very stable evolution (Figure 1). Its semi-major axis has only
short period oscillations with small amplitudes, while the eccentricity and
inclination show periodic secular evolution with amplitudes of $\Delta e \simeq 0.1$
and $\Delta I \simeq 3^\circ$. That is the expected evolution due to the 
secular perturbation caused by Jupiter.
Also in Figure 1 (in red) is shown the evolution of Sylvia's orbital elements
under Jupiter's perturbation, considering only the secular terms of the three-body 
disturbing function (chapter 7 of Murray \& Dermott 1999).

Now we present the results of the numerical integrations for a set of different dynamical systems.
Figure 2 shows the temporal evolution of the eccentricity ($e$), the inclination ($I$)
and the longitude of the ascending node ($\Omega$) of Romulus (left column) and Remus (right column). 
In each figure the results from three different systems are plotted:
the 3-body problem, Sylvia-Romulus-Remus (in red);
the 4-body problem, Sylvia-Romulus-Remus-Sun (in green);
the 5-body problem, Sylvia-Romulus-Remus-Sun-Jupiter (in blue).

\begin{table*}
\caption{Physical and orbital data of Sylvia's system (Marchis et al. 2005).}             
\label{table:1}      
\centering                          
\begin{tabular}{c c c c c c c c c c}        
\hline\hline                 
           & Mass (Kg)         & $a$     &  $e$ & $I (^\circ)$ &$\Omega (^\circ)$&$\omega (^\circ)$&$f (^\circ)$   &Orbital Period  &\\
\hline                        
   Sylvia  & $1.4780\times 10^{19}$ & 3.49 AU & 0.08 & 10.855   &266,195 & 73,342 & 8,51412 &  6.52 years &\\
   Romulus  & $3.6625\times 10^{15}$ & 1356 Km & 0.001& 1.7      &  273   & 101    & 81.88  &  3.65 days &\\
   Remus    & $2.1540\times 10^{14}$ &  706 Km & 0.016& 2.0      &  314   & 97     & 12.695 &  1.38 days &\\
  \hline                                   
\end{tabular}
\end{table*}

\begin{figure*}
\begin{center}
\includegraphics[height=6.0cm]{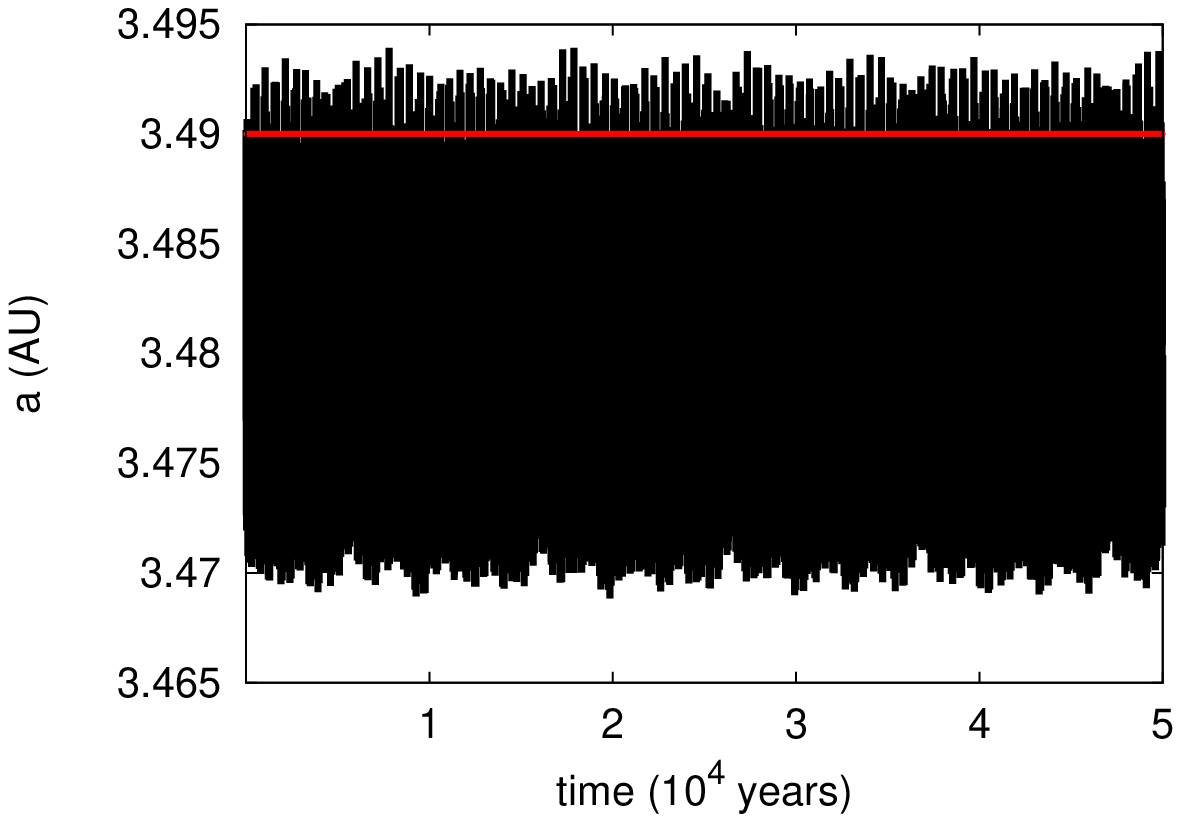}
\includegraphics[height=6.0cm]{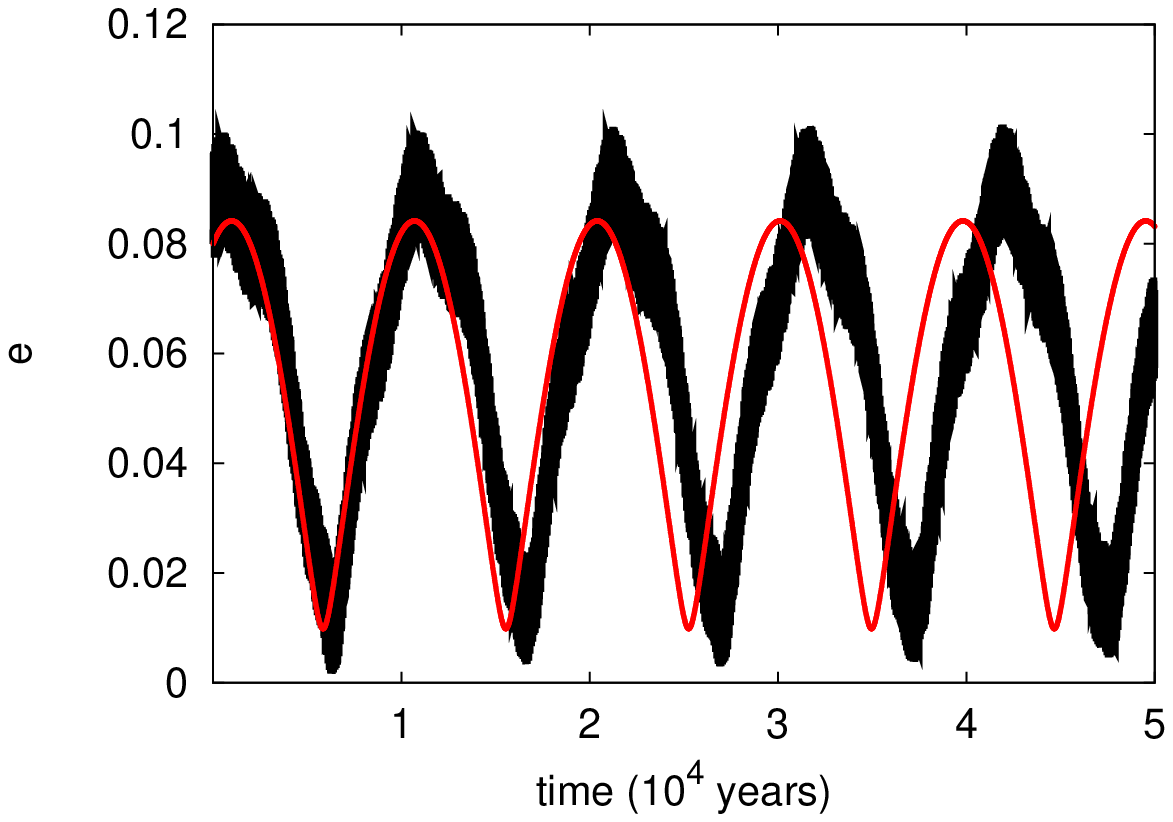}
\includegraphics[height=6.0cm]{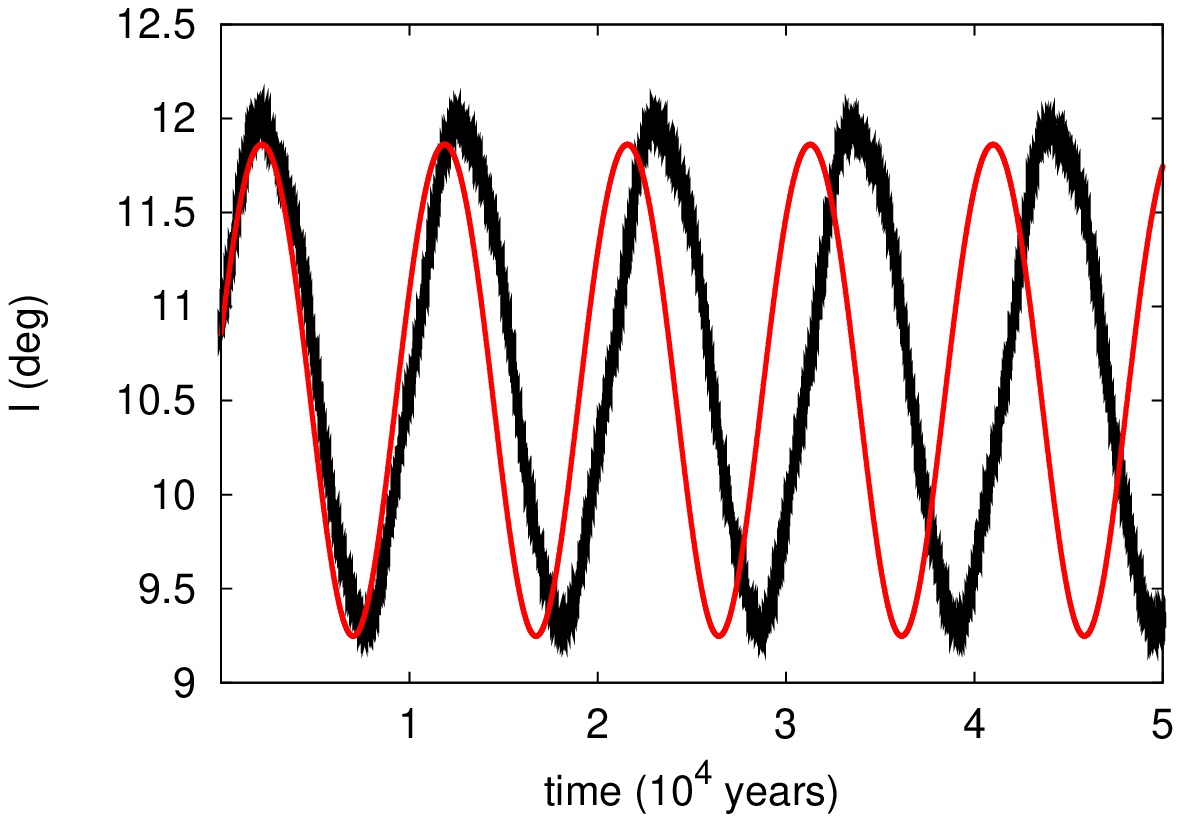}
\includegraphics[height=6.0cm]{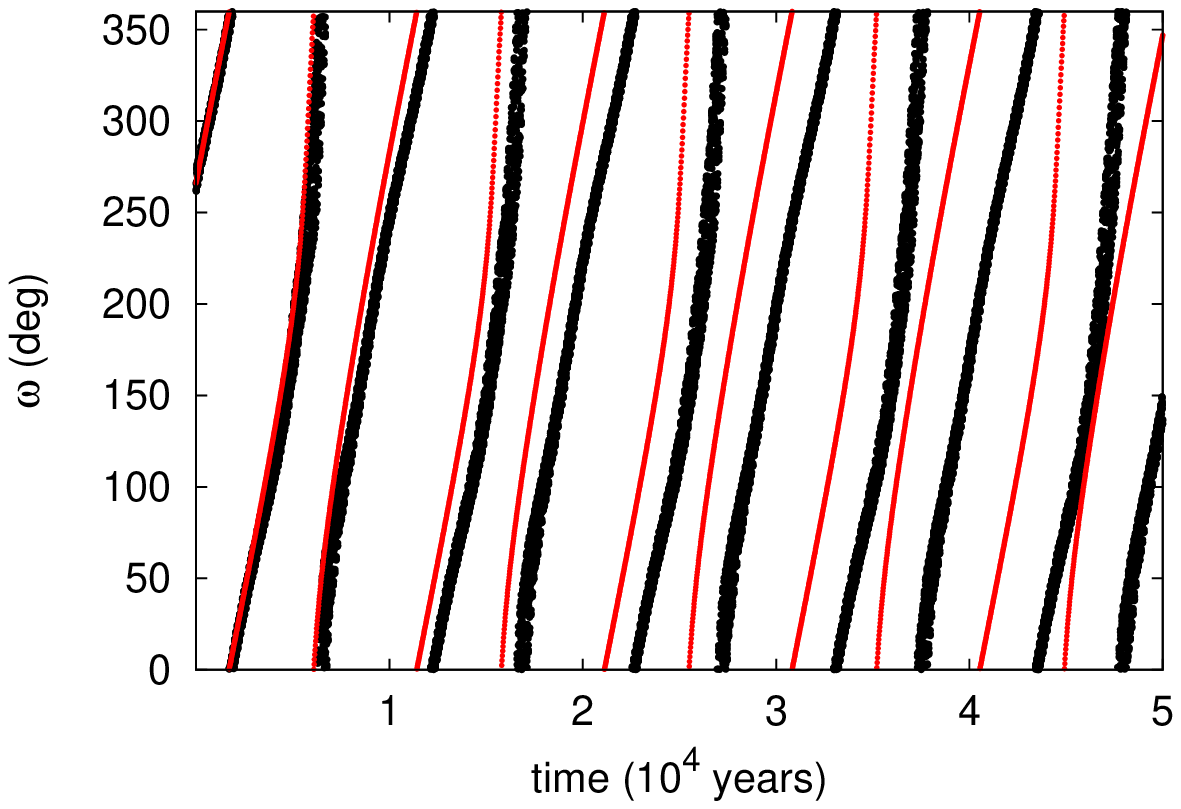}
\includegraphics[height=6.0cm]{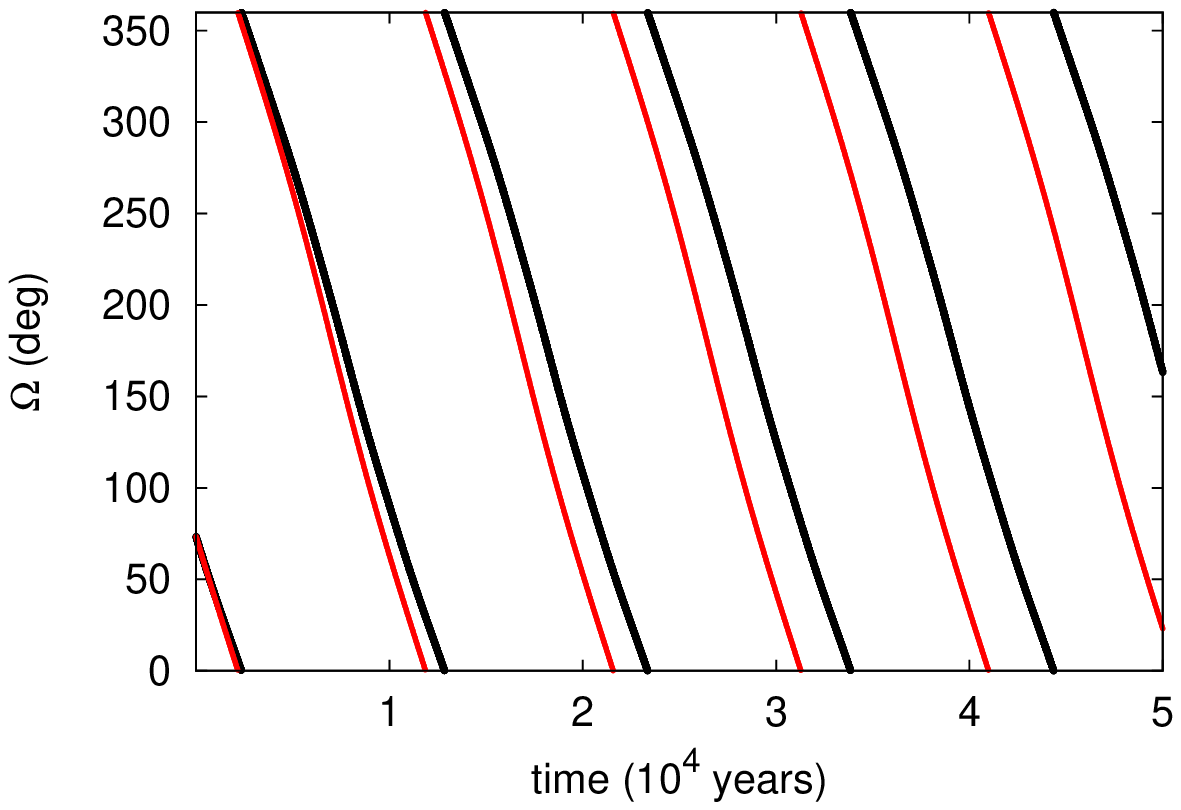}
\end{center}
\caption{\label{figura5}: Temporal evolution of the semi-major axis, eccentricity, inclination, argument of pericentre and longitude of the ascending node of Sylvia. These are the results from the numerical integration of the 5-body problem - Sylvia-Romulus-Remus-Sun-Jupiter (in black) and from secular perturbation theory - Sylvia-Sun-Jupiter (in red).}
\end{figure*}

\begin{figure*}
\begin{center}
\includegraphics[height=6.0cm]{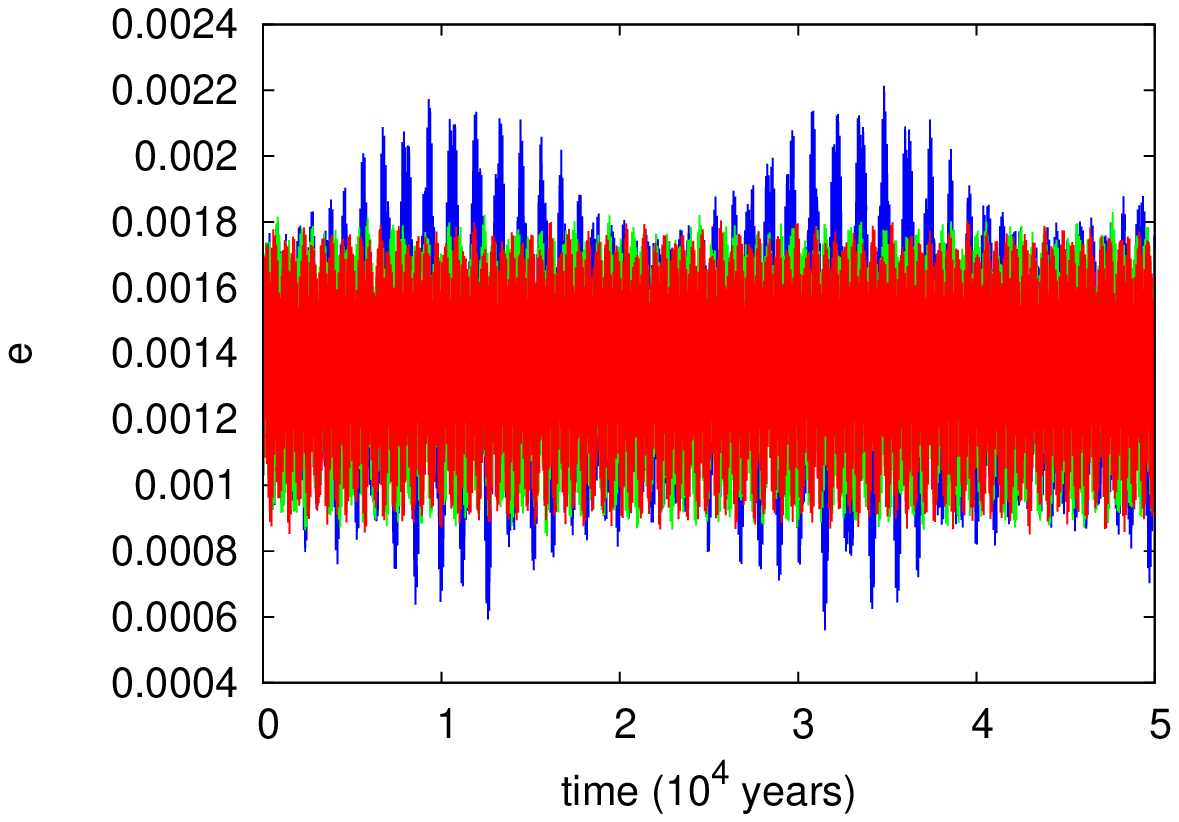}
\includegraphics[height=6.0cm]{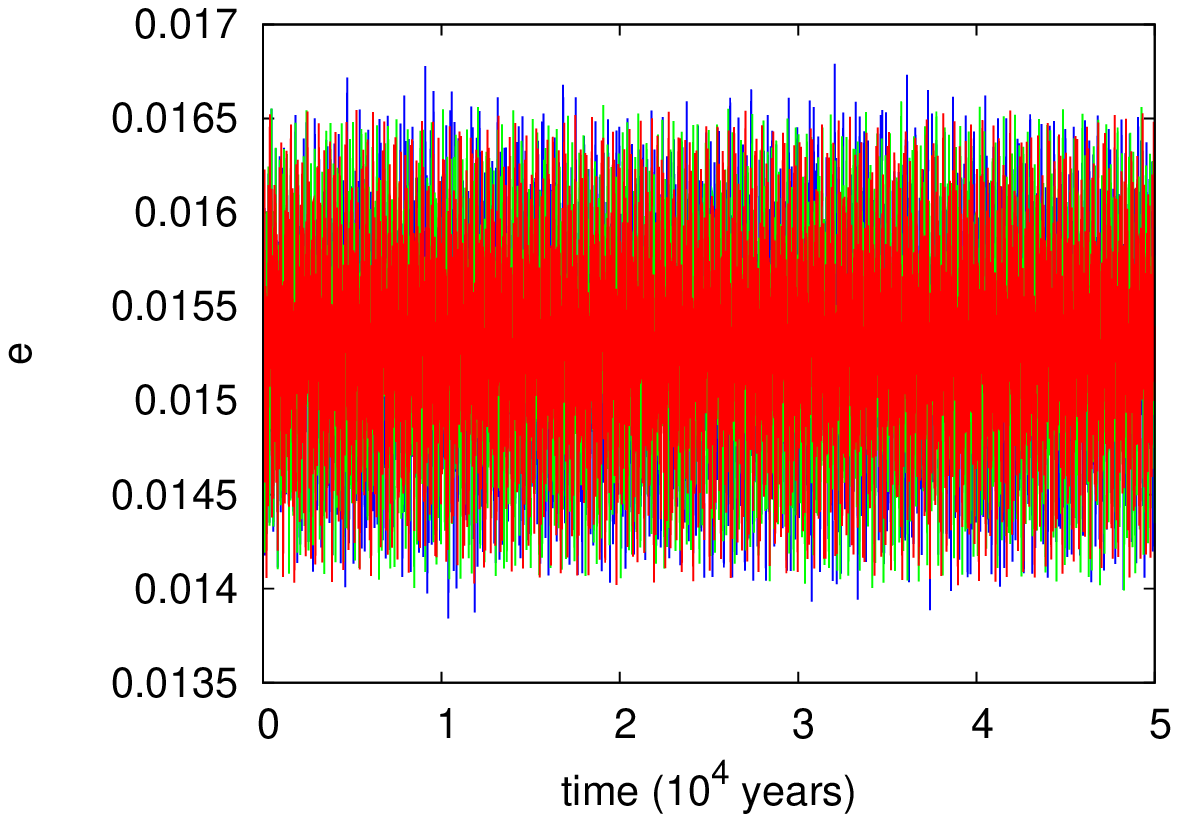}
\includegraphics[height=6.0cm]{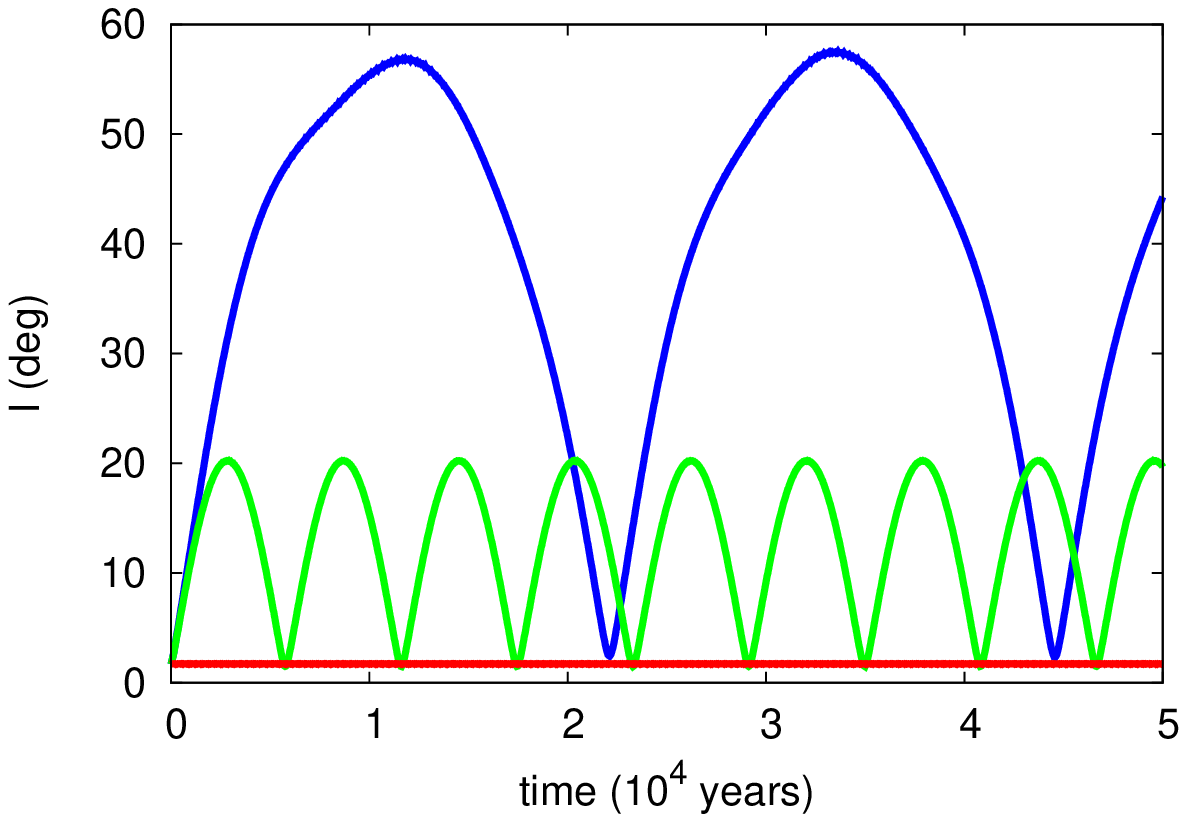}
\includegraphics[height=6.0cm]{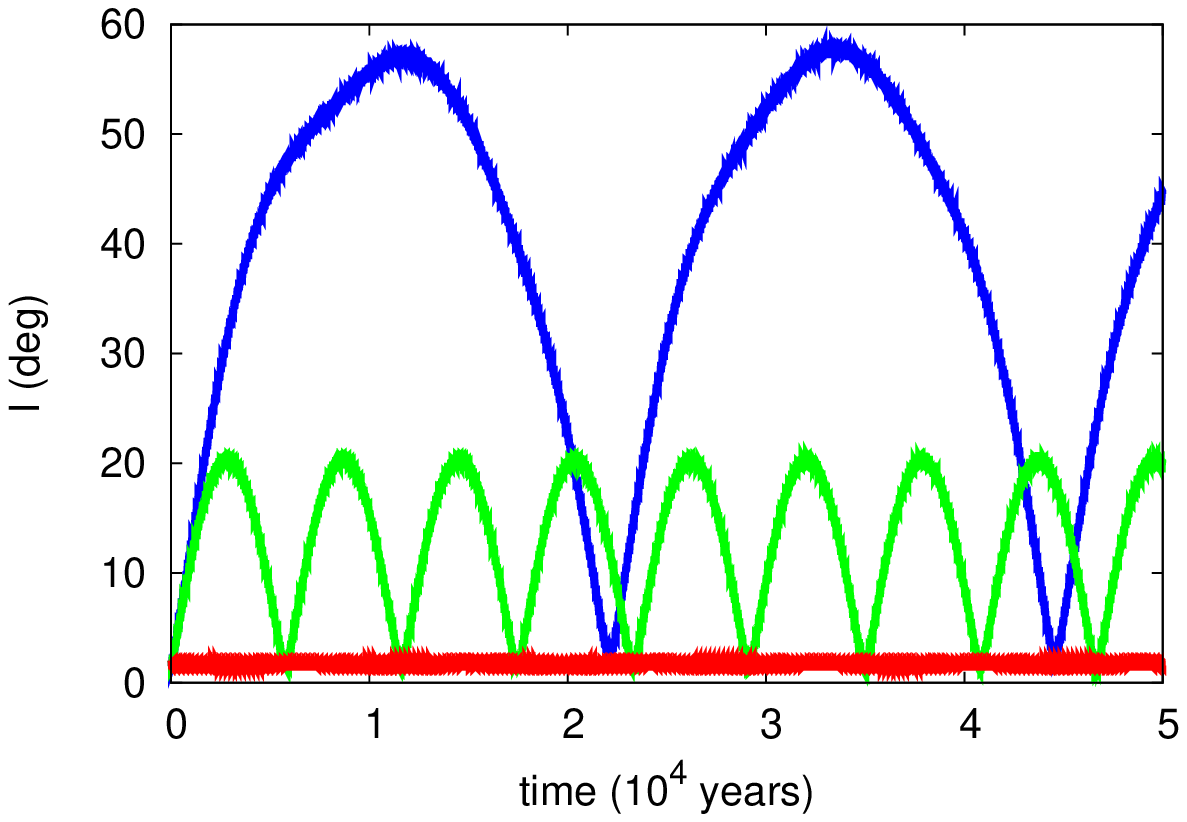}
\includegraphics[height=6.0cm]{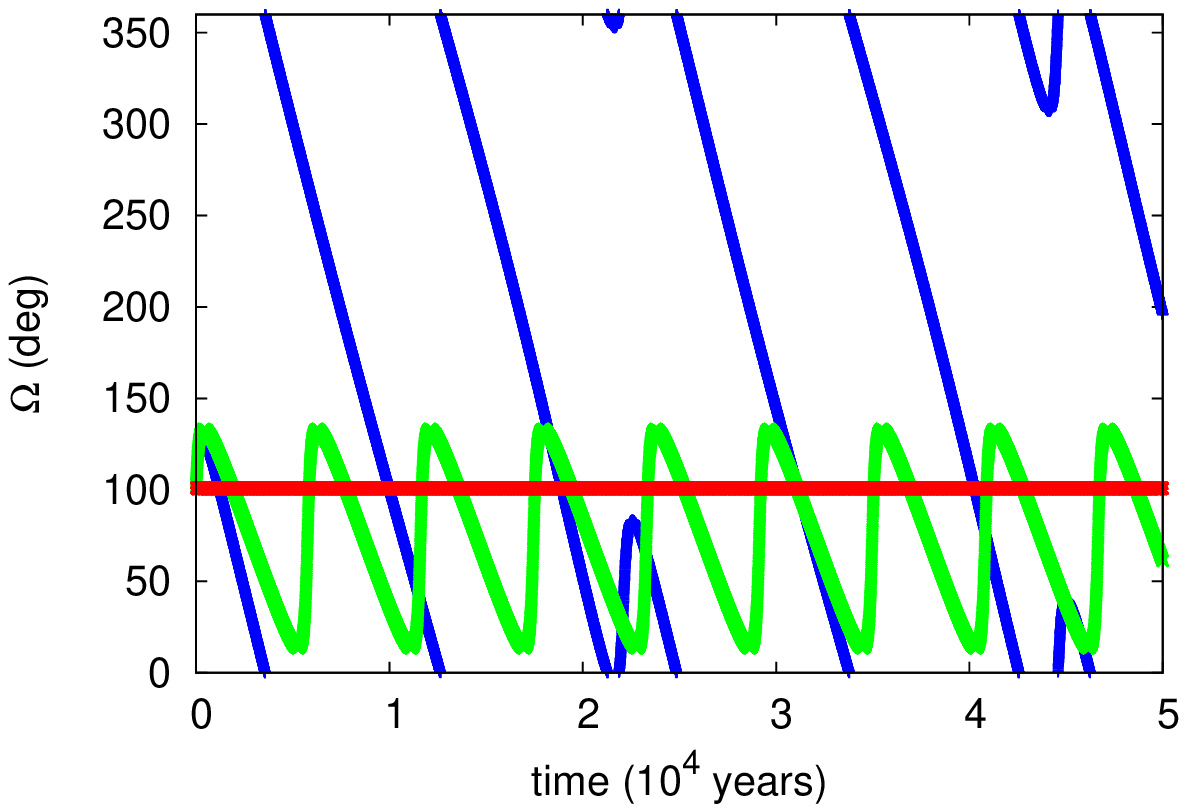}
\includegraphics[height=6.0cm]{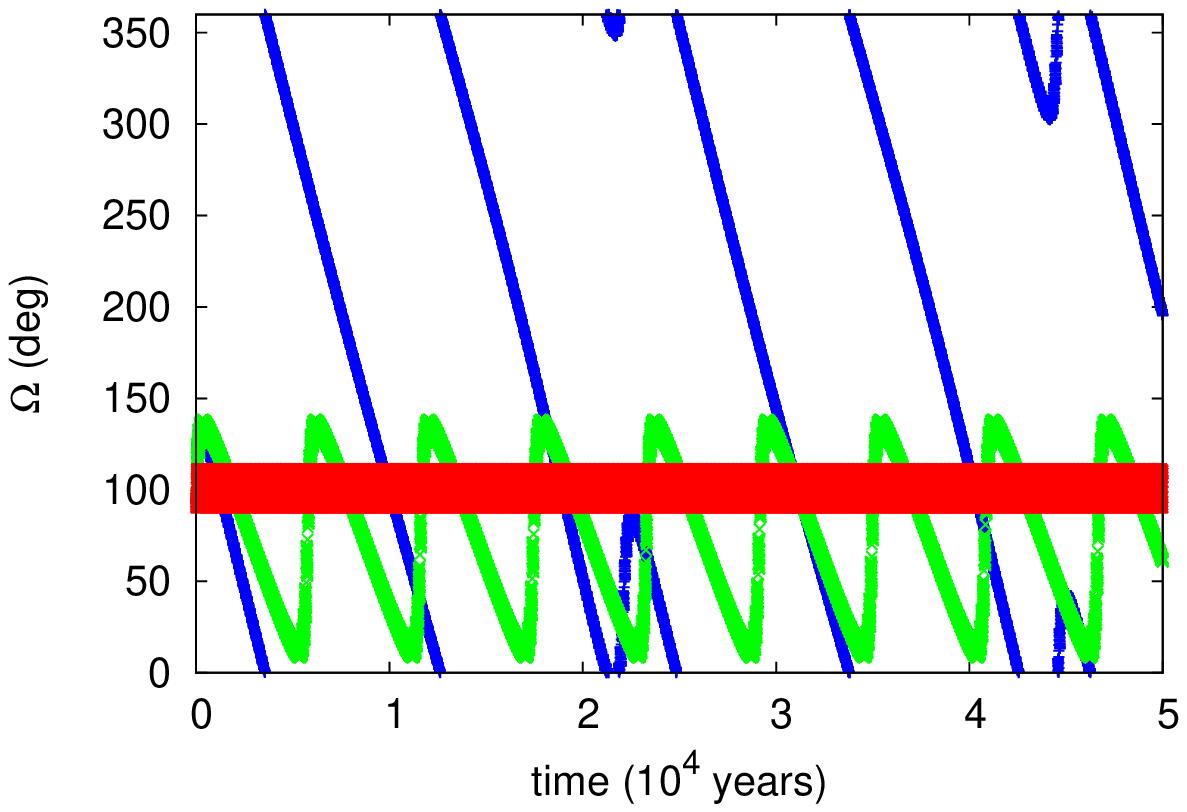}
\end{center}
\caption{\label{figura1}: Temporal evolution of the eccentricity, the inclination and the longitude of the ascending node of Romulus (left column) and Remus (right column). In each figure are plotted the results from three different numerical integrations. 1) In red, the 3-body problem - Sylvia-Romulus-Remus. 2) In green, the 4-body problem - Sylvia-Romulus-Remus-Sun. 3) In blue, the 5-body problem - Sylvia-Romulus-Remus-Sun-Jupiter.}
\end{figure*}

The results from the 3-body problem simulation, Sylvia-Romulus-Remus, show no significant
variation of the satellite's orbital elements. 
The inclusions of the Sun and Jupiter do not affect
the evolution of the eccentricities of the satellites.
However, their orbital inclinations show a significant change.
They present a periodic secular evolution, having amplitude of about $20^\circ$
with period of almost $6\times 10^3$ years (only Sun),
and amplitude of about $58^\circ$  with period of almost $22\times 10^3$ years (Sun and Jupiter). 
Their longitudes of the node either librate with amplitude of more than $120^\circ$
(only Sun) or circulate (Sun + Jupiter).

We also note that the evolution of $i$ and $\Omega$ for both satellites present very similar amplitudes and frequencies. Of course, these evolutions are mainly due to the perturbations of the Sun (green) and Sun and Jupiter (blue), respectively. 
However, the fact of them being so similar is associated to a connection between Romulus and Remus, 
which will be discussed in the following section.

\section{The Romulus-Remus Connection}

The gravitational interaction between Romulus and Remus produces different outcomes 
according to the dynamical system considered. 
In this section we concentrate the analysis on the temporal evolution of the satellites' orbital inclinations.
We compare the results from our numerical integrations with results from the secular perturbation theory (see for instance Murray \& Dermott (1999)).

From the numerical integrations of the 3-body system, Sylvia-Romulus-Remus (Figure 2, in red),
we see that Romulus produces oscillations of small amplitude ($<4^\circ$) on Remus's inclination,
which in return produces an even smaller amplitude of oscillation on Romulus's inclination ($<1^\circ$).
These results are very close to those obtained from the secular perturbation theory (Figure 3, first row).
The direct coupling between Romulus and Remus, showing that when the inclination of one increases the other decreases, is clearly seen in the zoom of these plots (Figure 3, first row).

The secular perturbation from the Sun on Romulus and on Remus, separately, produces oscillations with
the same amplitude ($\sim 20^\circ$), but with different periods (Figure 3, second row).
The period of oscillation for the inclination of Remus is more than twice that for Romulus.
However, when the secular perturbation from the Sun and one of the satellites on the other
satellite is computed, the amplitude and also the period of the oscillation of the inclination
of both satellites are almost the same (Figure 3, third row). Actually, the gravitational interaction between Romulus and Remus is such that Remus's inclination follows very close the behaviour of Romulus's inclination.  
Such results are in very good agreement with our numerical integrations of the 4-body problem - Sylvia-Romulus-Remus-Sun (Figure 3, last row).

This connection between Romulus and Remus is due to a secular resonance. 
The longitude of the ascending node of both satellites are librating (Figure 2, last row, in green).
The evolution of the resonant angle, given by  $\phi=\Omega_{\rm Rom} -\Omega_{\rm Rem}$, shows 
a libration of very short amplitude (Figure 4, middle), i.e., the satellites' longitudes of the ascending nodes  get locked and librate with almost the same frequency, $\dot\Omega_{\rm Rom} \approx  \dot\Omega_{\rm Rem}$. This result is also verified by the results from the secular perturbation theory (Figure 4, left).

Despite of the fact that in the 5-body problem, Sylvia-Romulus-Remus-Sun-Jupiter, the longitude of the ascending node of both satellites are circulating (Figure 2, last row, in blue), that same secular resonance occurs  (Figure 4, right).
Therefore, the evolutions of the orbital inclination and longitude of the node of the two satellites are very much similar in the 5-body system due to this secular resonance.
If the gravitational influence of one of the satellites is not taken into account their orbital evolution are very different from each other (Figure 5). The amplitude of Remus's inclination is smaller than that of Romulus and their eccentricities grow erratically. 
So, we conclude that the direct gravitational perturbation of one satellite on the other does not produce any significant variation on their orbital elements. However, when these satellites are under strong perturbations from the Sun and Jupiter, the two satellites get locked in a secular resonance such that the orbital evolution of one is very similar to the other.

\begin{figure*}
\begin{center}
\includegraphics[height=5.0cm]{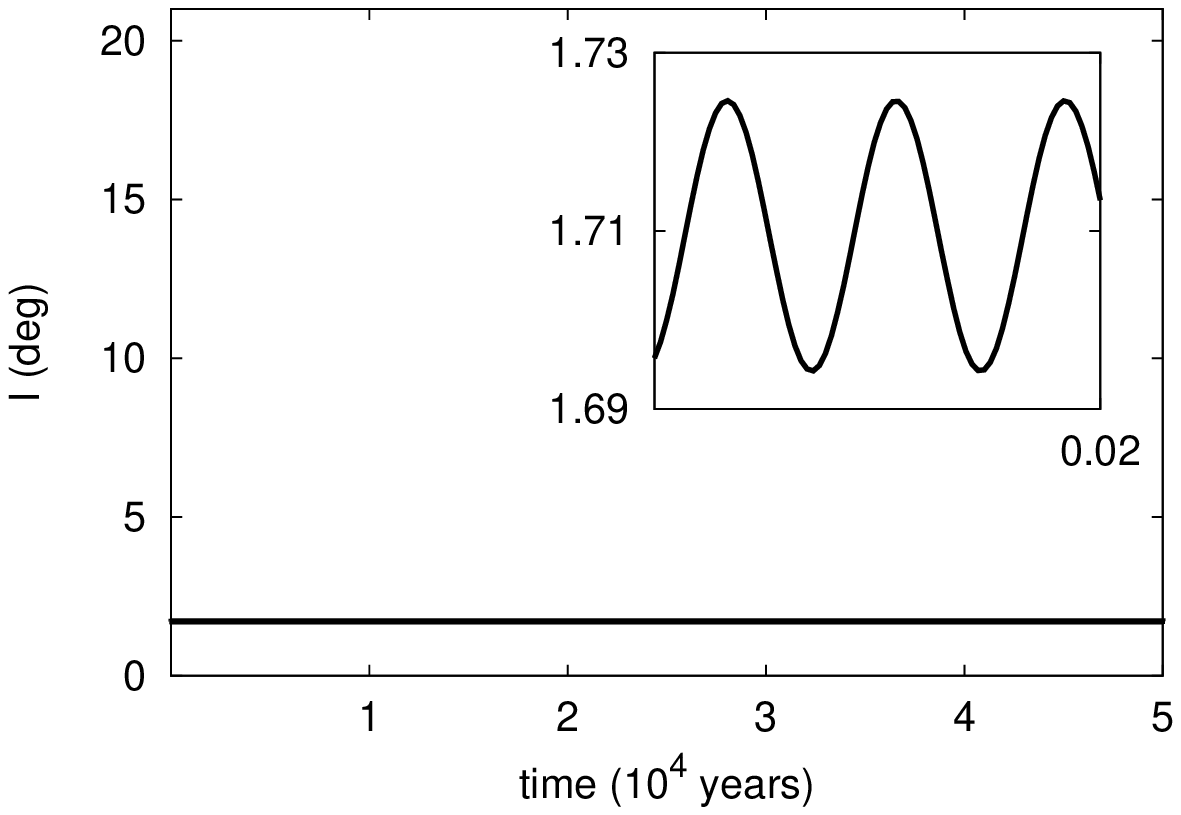}
\includegraphics[height=5.0cm]{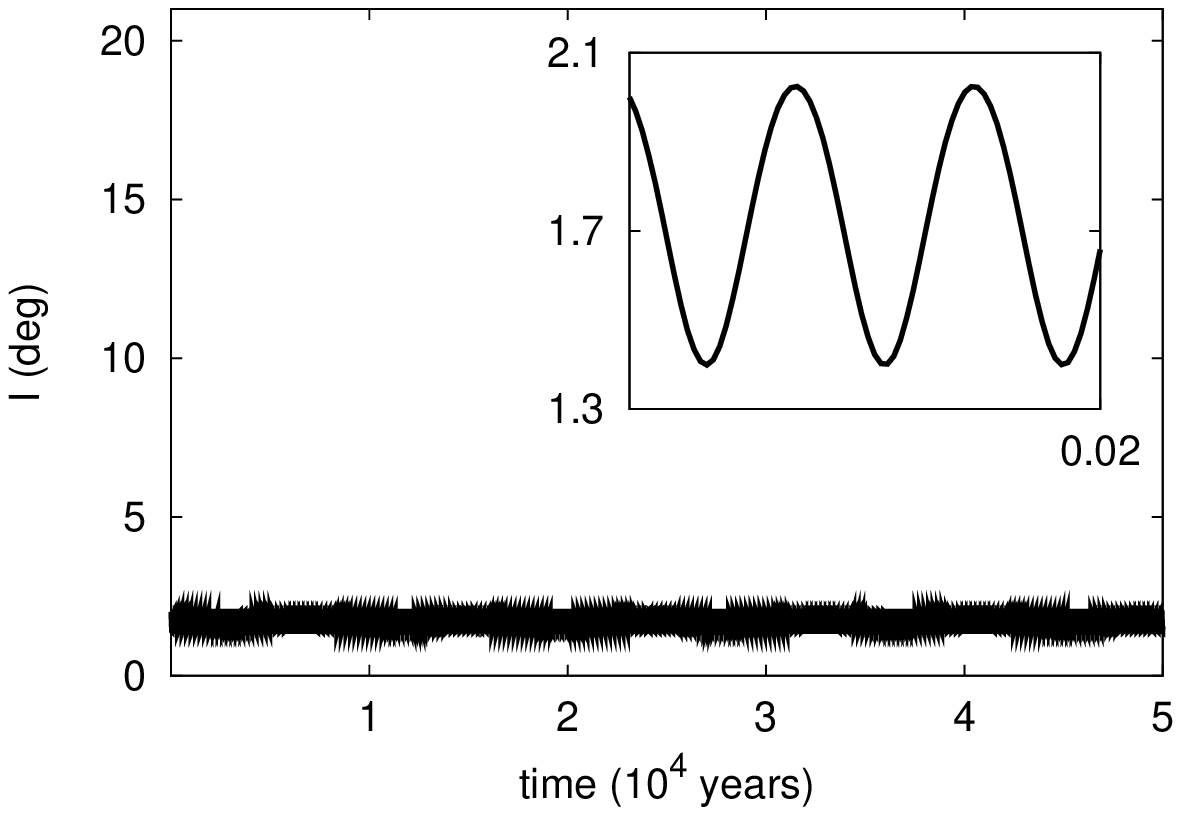}
\includegraphics[height=5.0cm]{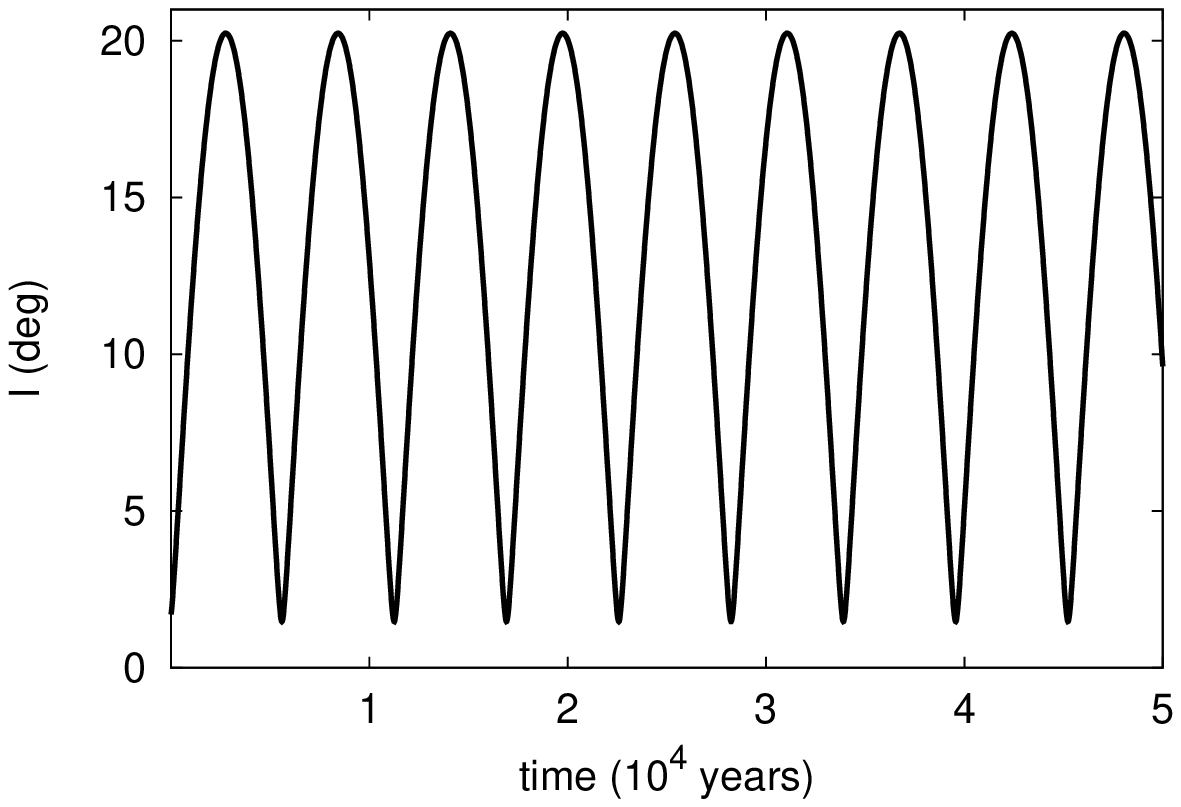}
\includegraphics[height=5.0cm]{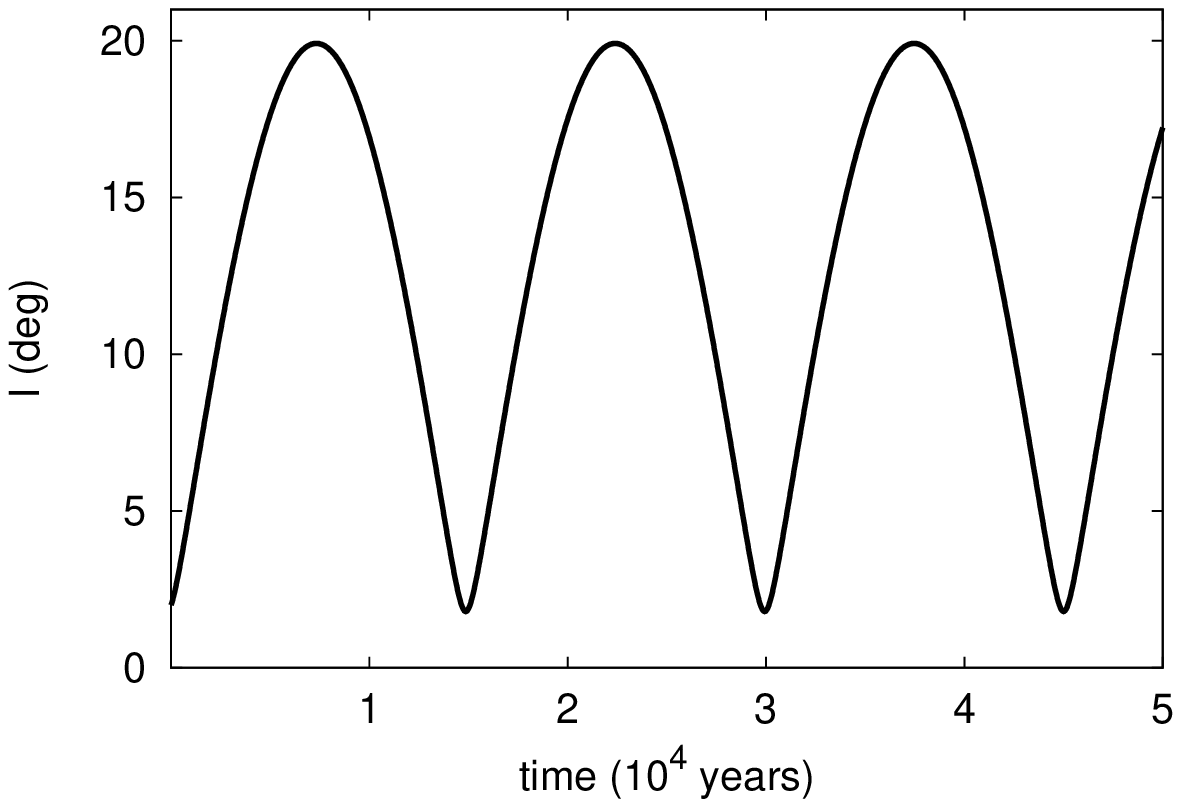}
\includegraphics[height=5.0cm]{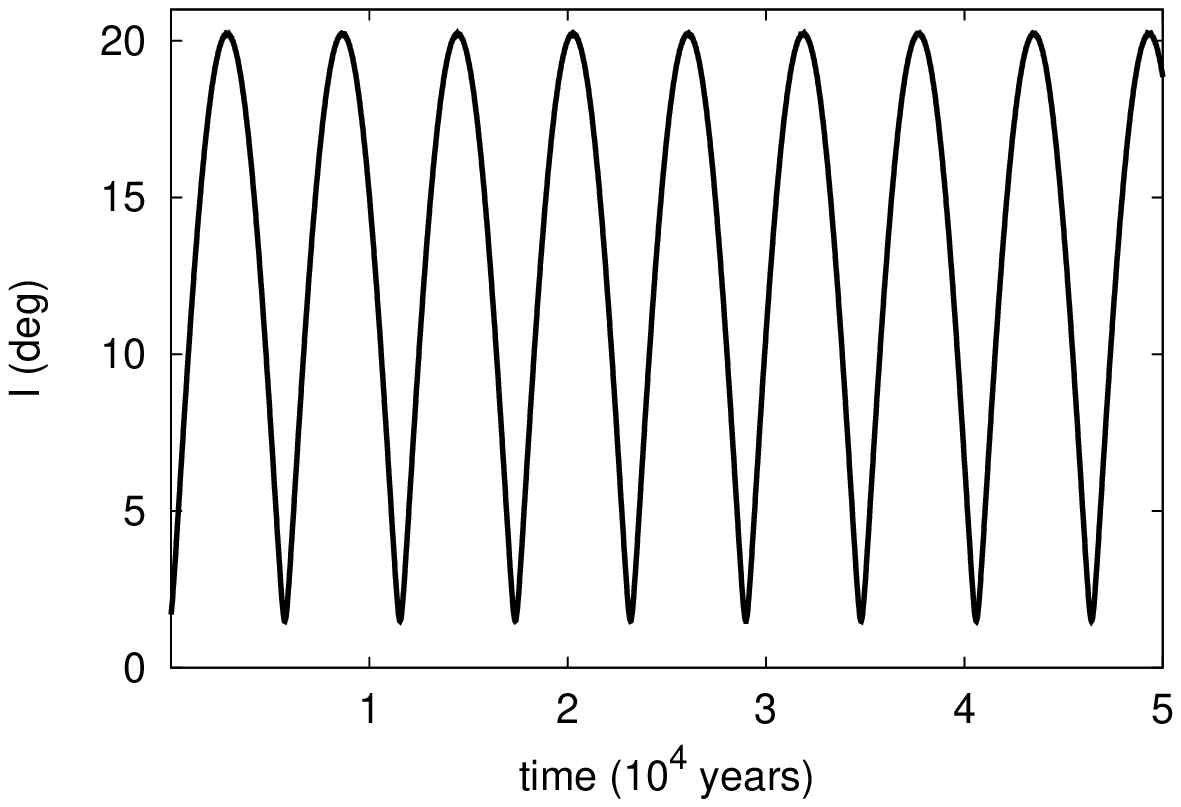}
\includegraphics[height=5.0cm]{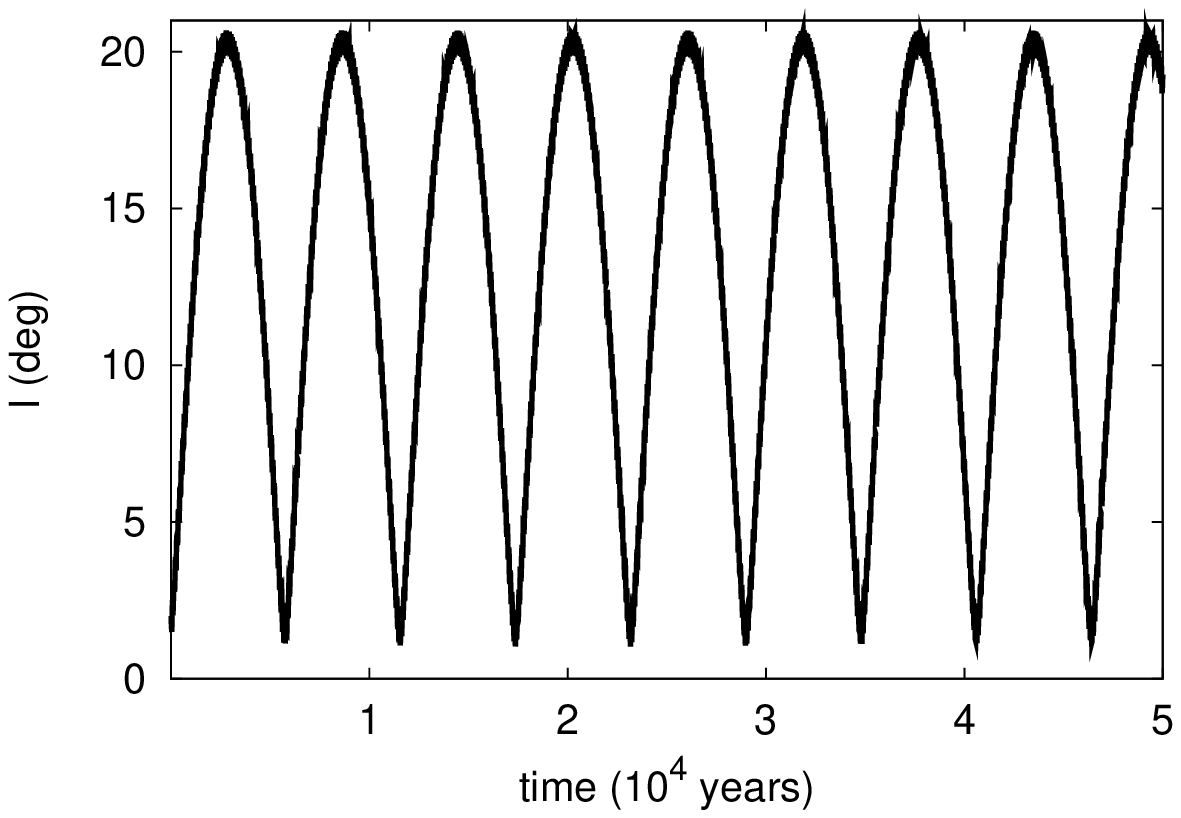}
\includegraphics[height=5.0cm]{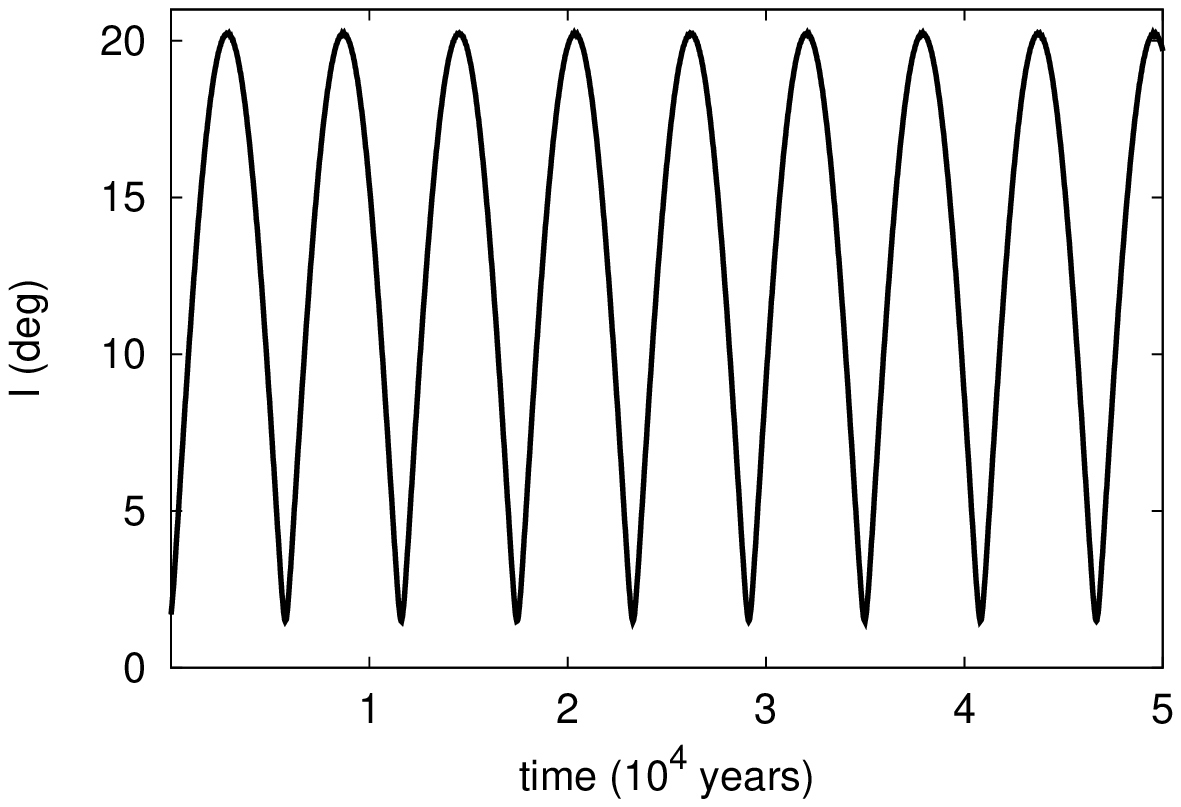}
\includegraphics[height=5.0cm]{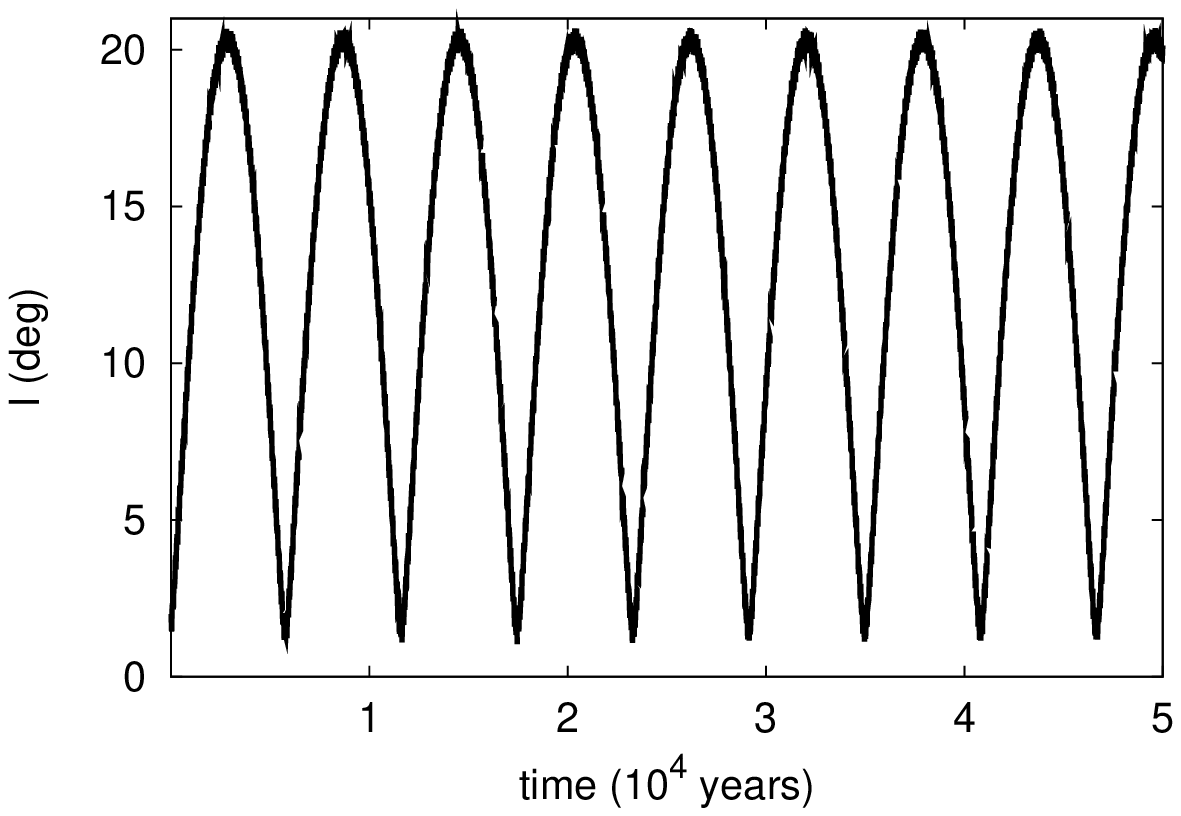}
\end{center}
\caption{\label{figura2}: Temporal evolution of the orbital inclinations of Romulus (left column) and Remus (right column). 
In the first row are presented the results from the secular perturbation theory for the 3-body system - Sylvia-Romulus-Remus. It is also shown a zoom of the plots in the top right corner of the respective figure.
In the second row are the results from the secular perturbation theory for the 3-body systems: Sylvia-Romulus-Sun (left) and Sylvia-Remus-Sun (right) 
In the third row are presented the results from the secular perturbation theory for the 4-body system - Sylvia-Romulus-Remus-Sun.
In the last row are presented the results from the numerical integration for the 4-body system - Sylvia-Romulus-Remus-Sun (reproduced from Figure 2, in green) 
}
\end{figure*}

\begin{figure*}
\begin{center}
\includegraphics[width=5.5cm]{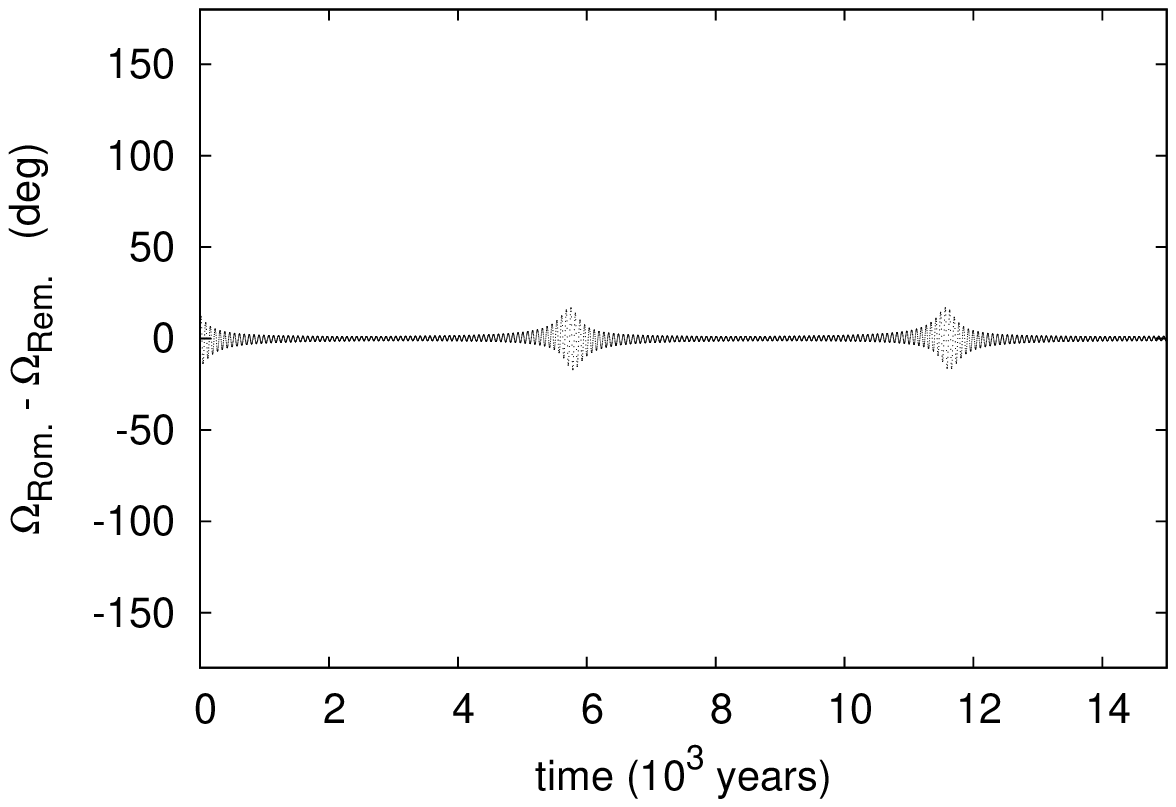}
\includegraphics[width=5.5cm]{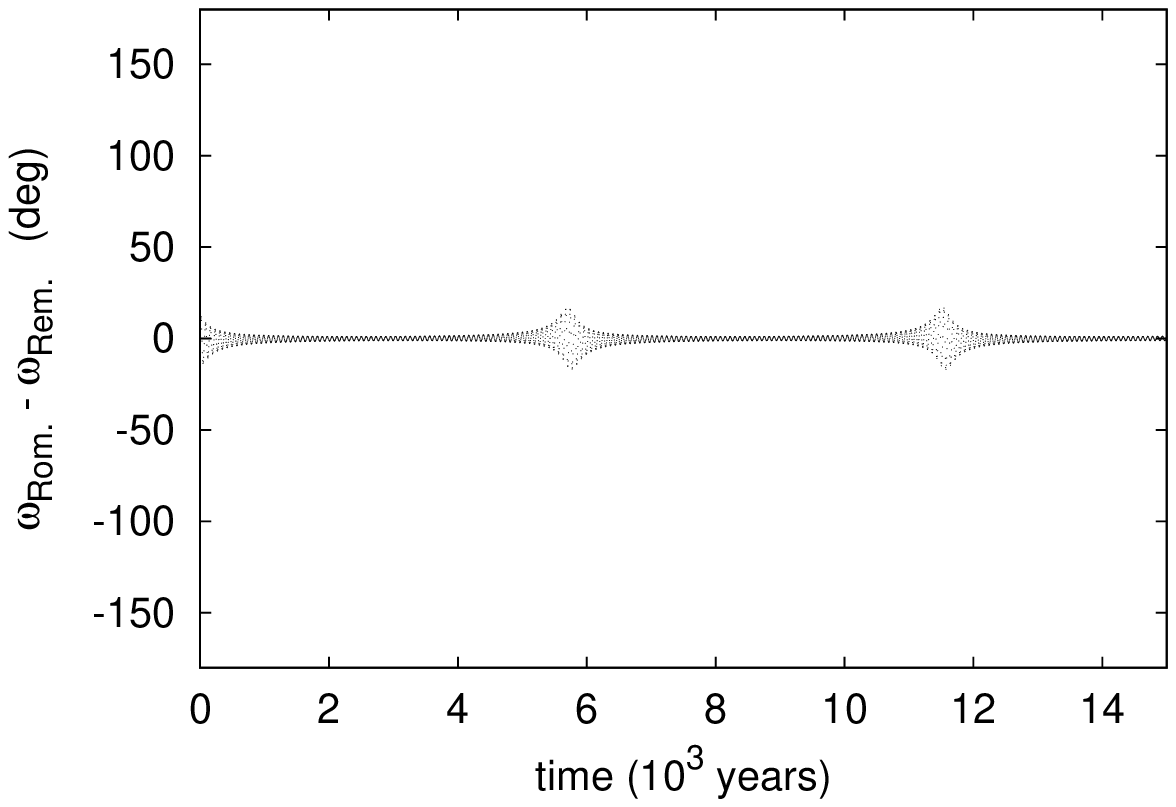}
\includegraphics[width=5.5cm]{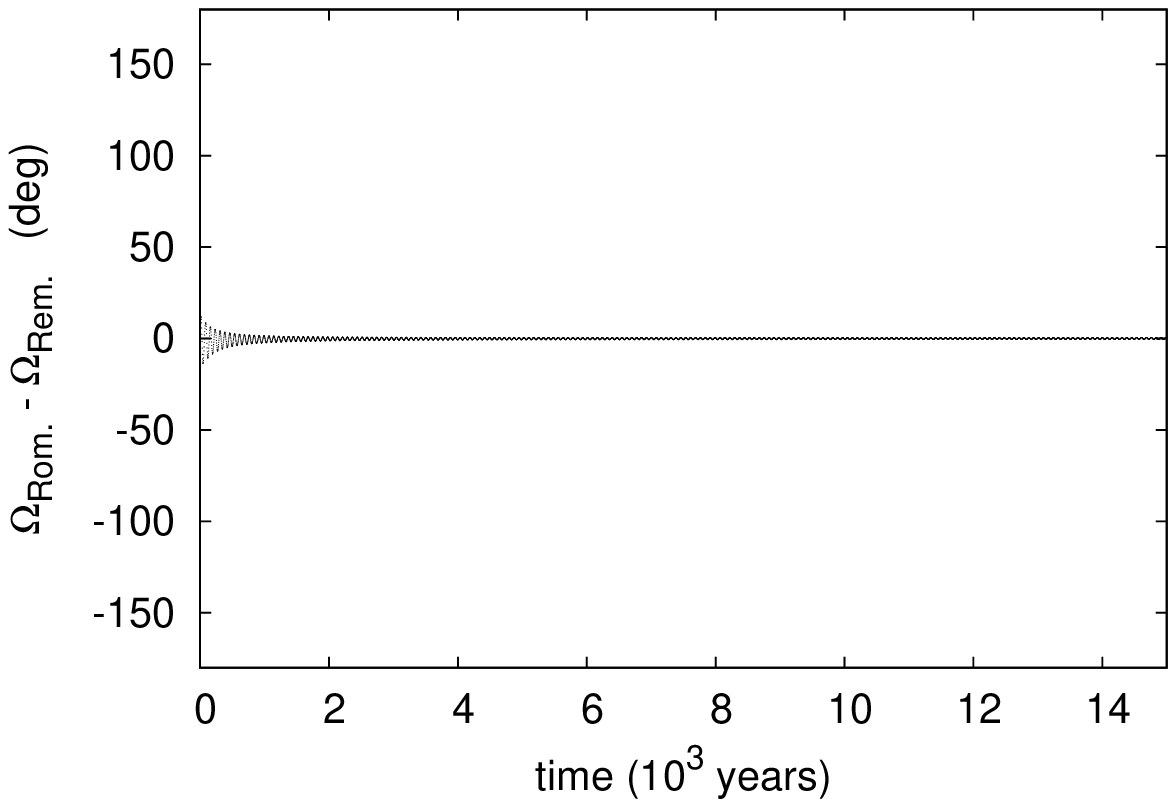}
\end{center}
\caption{\label{figura3}: Temporal evolution of the resonant angle, given by the difference between the longitudes of the ascending nodes of Romulus and Remus,  $\phi=\Omega_{\rm Rom} -\Omega_{\rm Rem}$. The left plot is the result from
the secular perturbation theory for the 4-body system - Sylvia-Romulus-Remus-Sun. The central plot is the result from the numerical integration for the 4-body problem - Sylvia-Romulus-Remus-Sun. The right plot is the result from the numerical integration for the 5-body problem - Sylvia-Romulus-Remus-Sun-Jupiter.
}
\end{figure*}

\begin{figure*}
\begin{center}
\includegraphics[height=5.0cm]{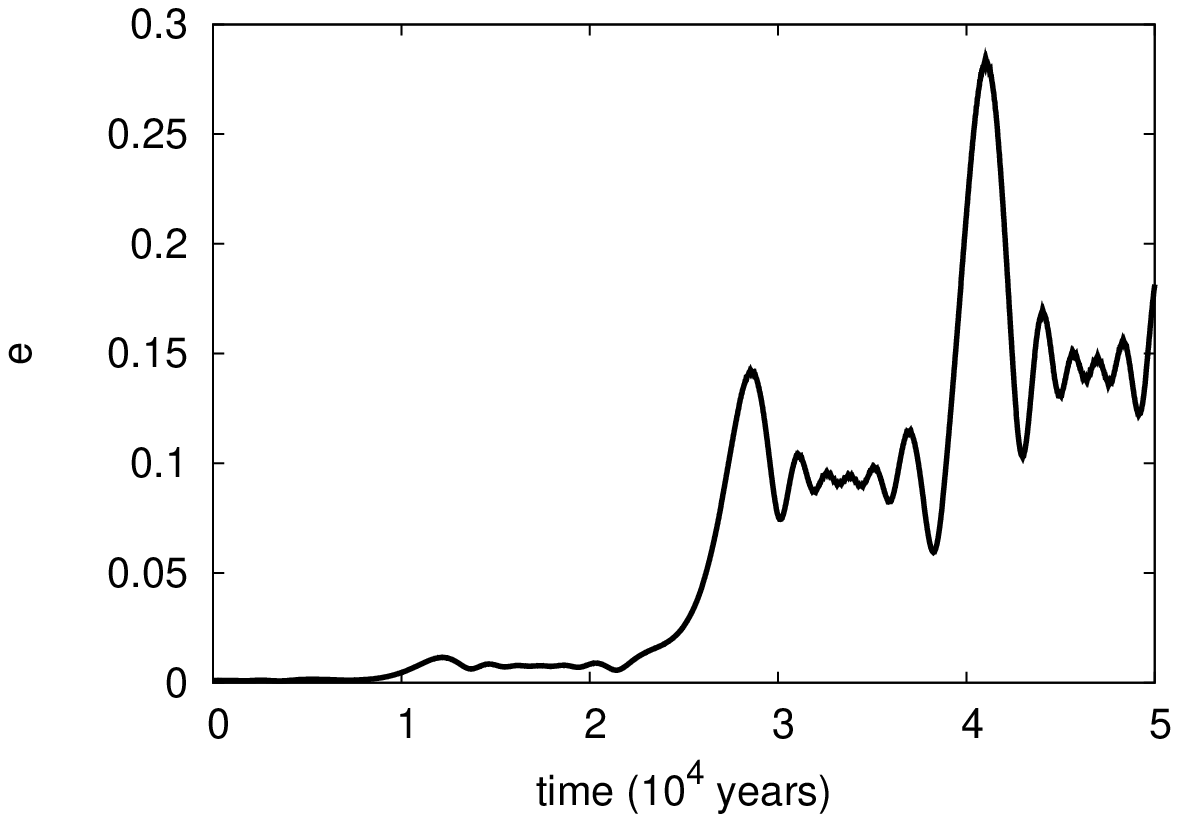}
\includegraphics[height=5.0cm]{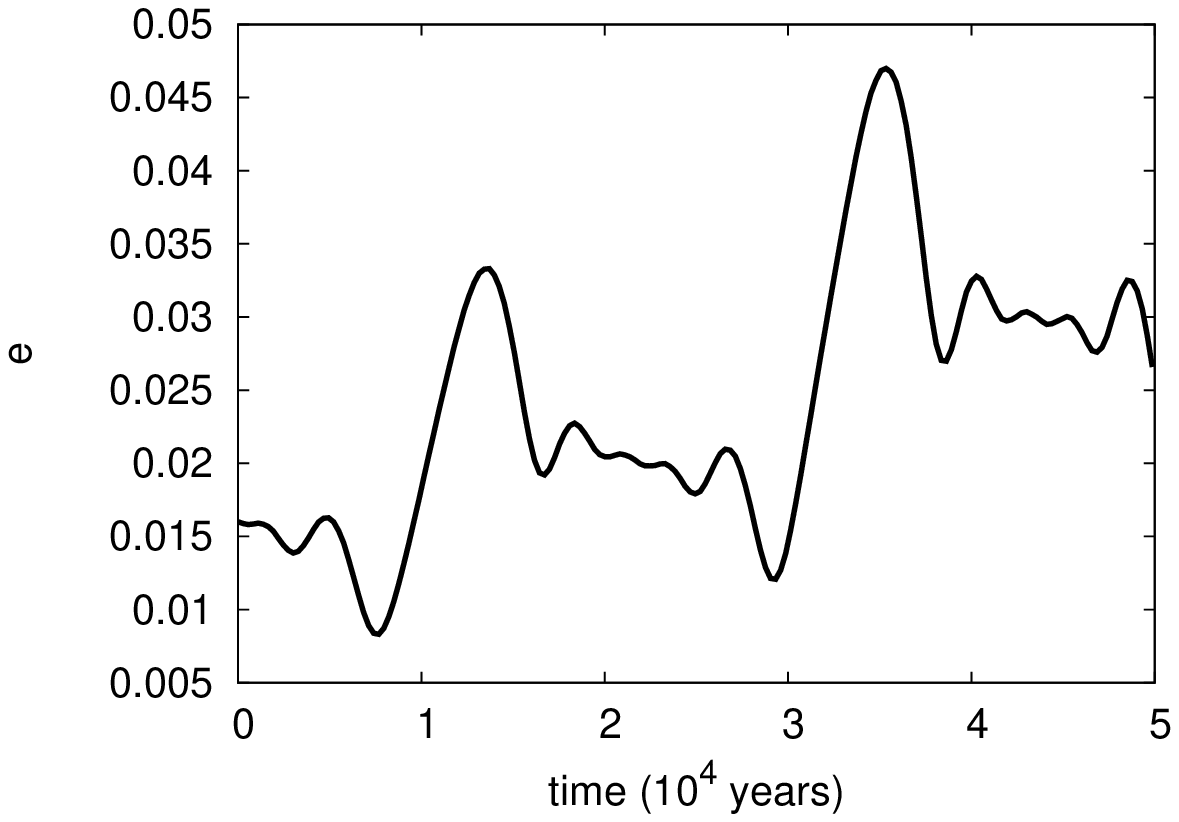}
\includegraphics[height=5.0cm]{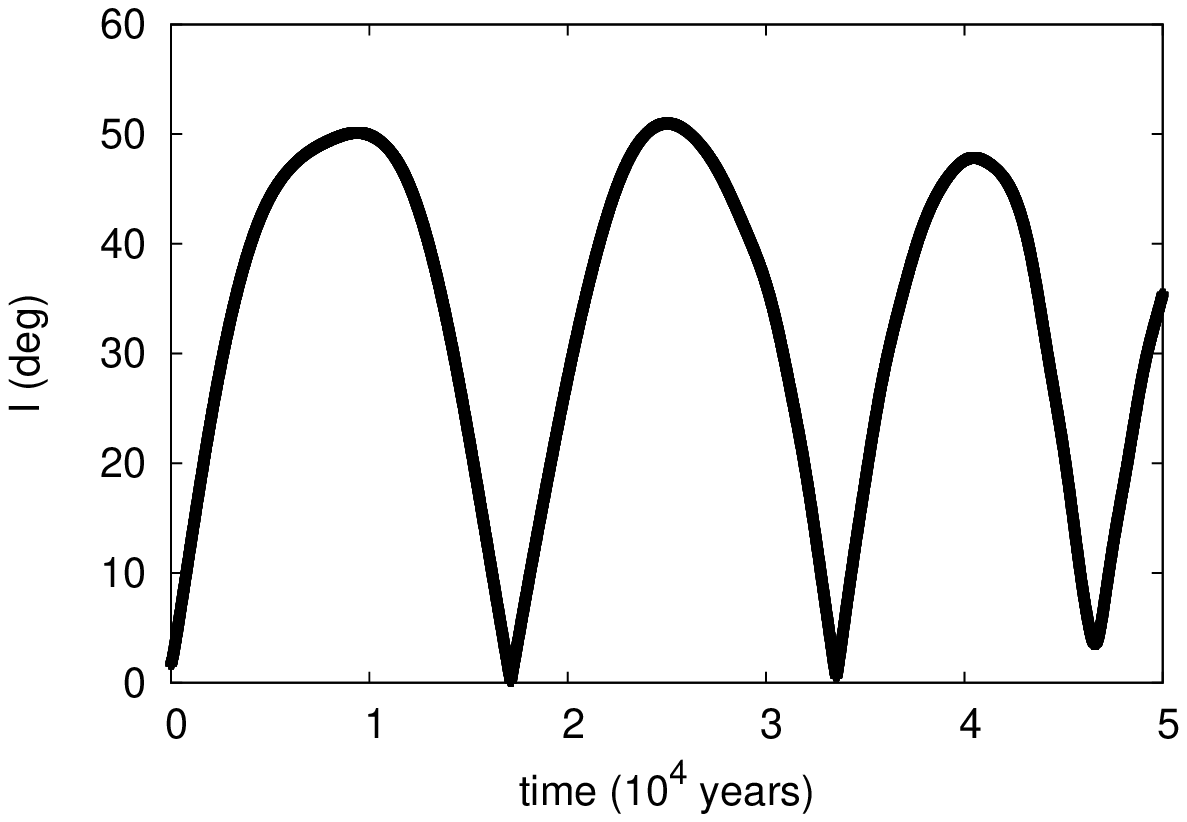}
\includegraphics[height=5.0cm]{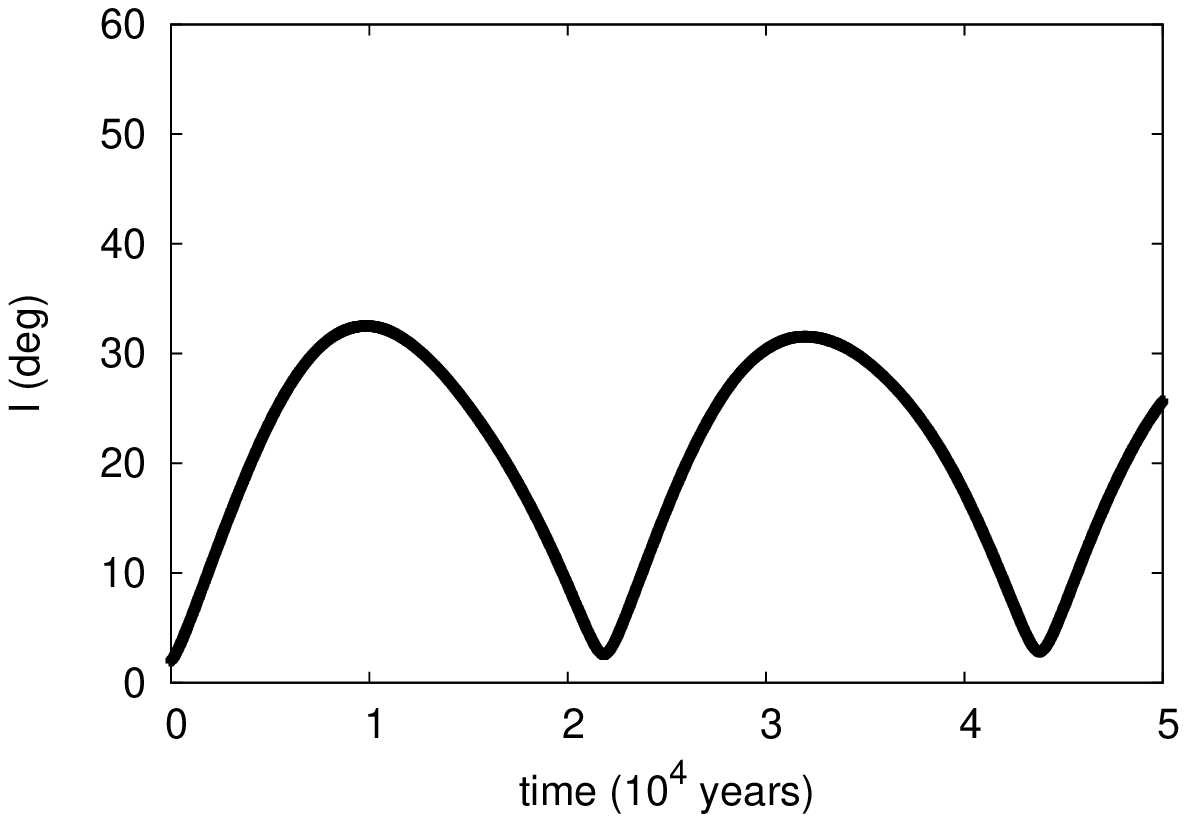}
\includegraphics[height=5.0cm]{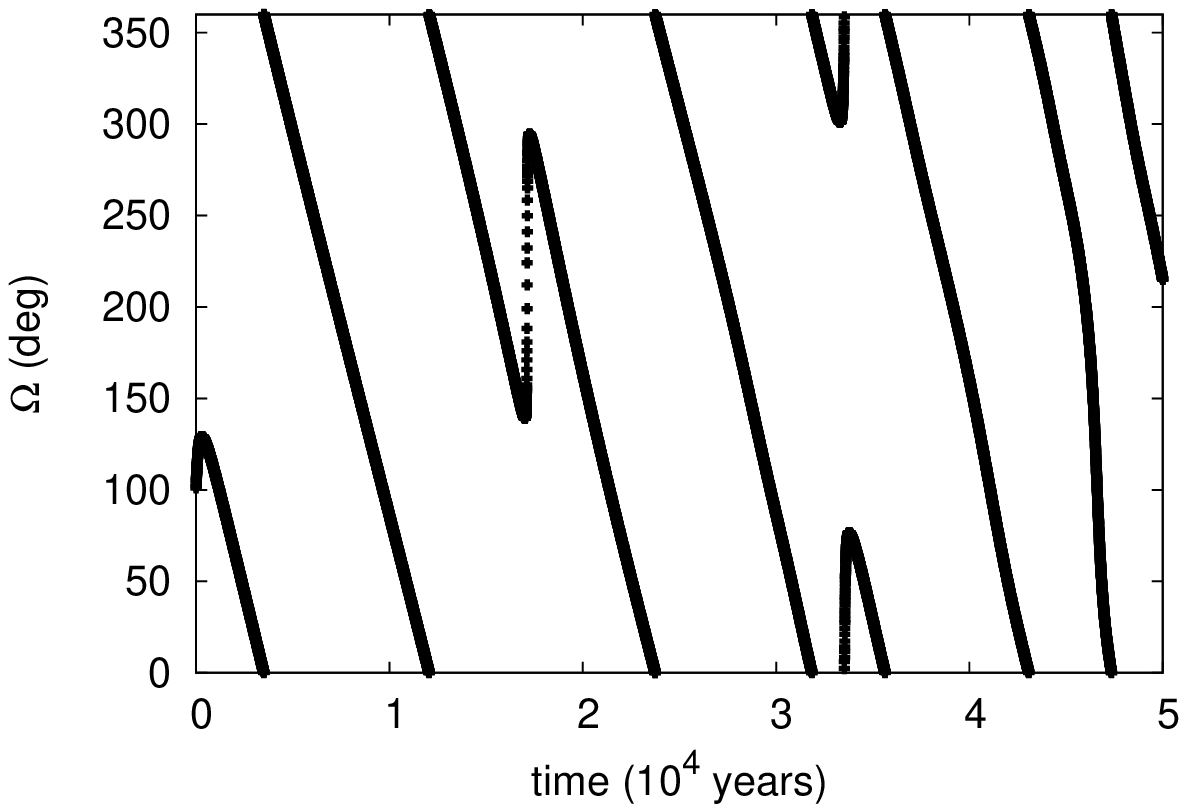}
\includegraphics[height=5.0cm]{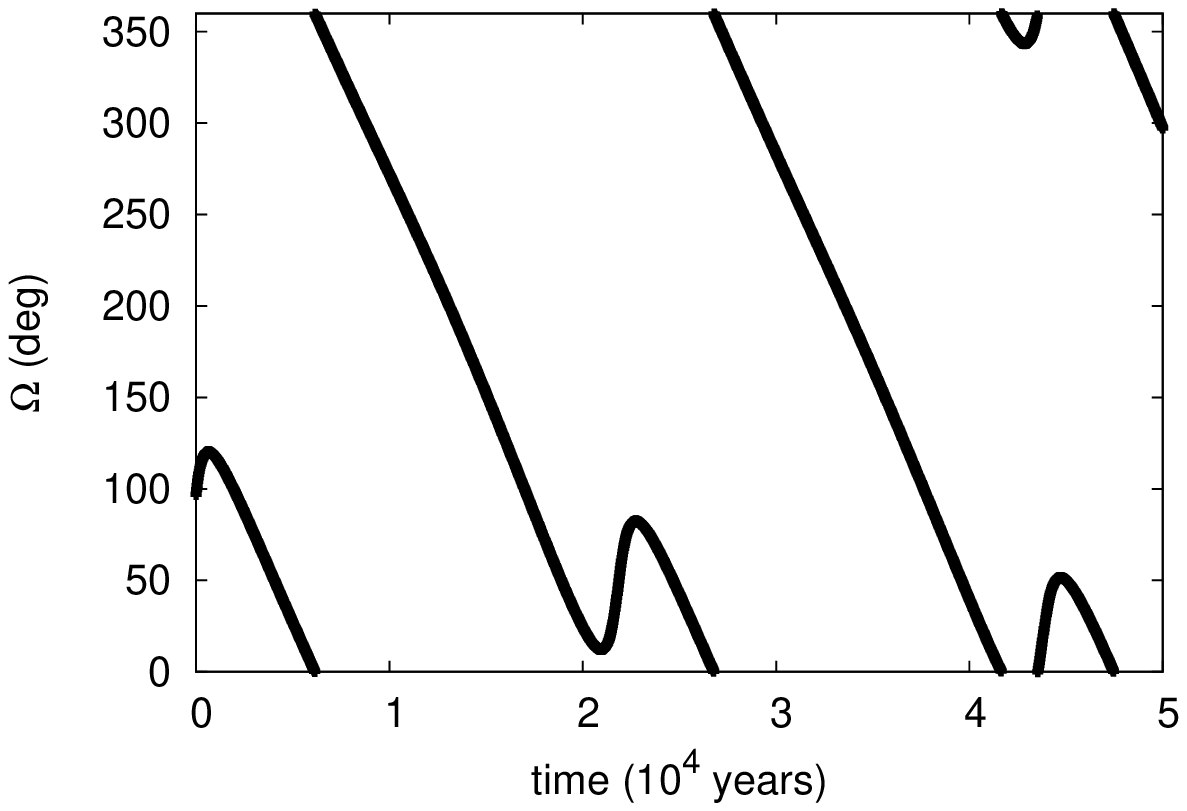}
\includegraphics[height=5.0cm]{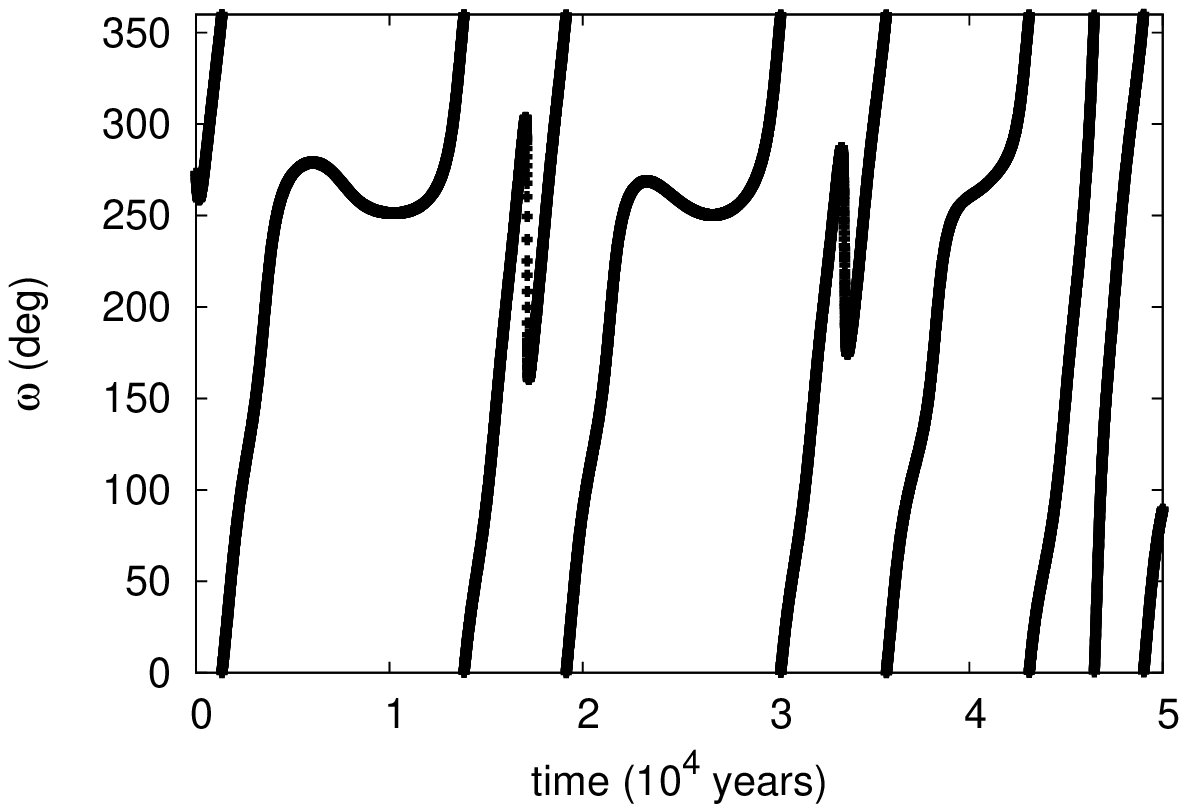}
\includegraphics[height=5.0cm]{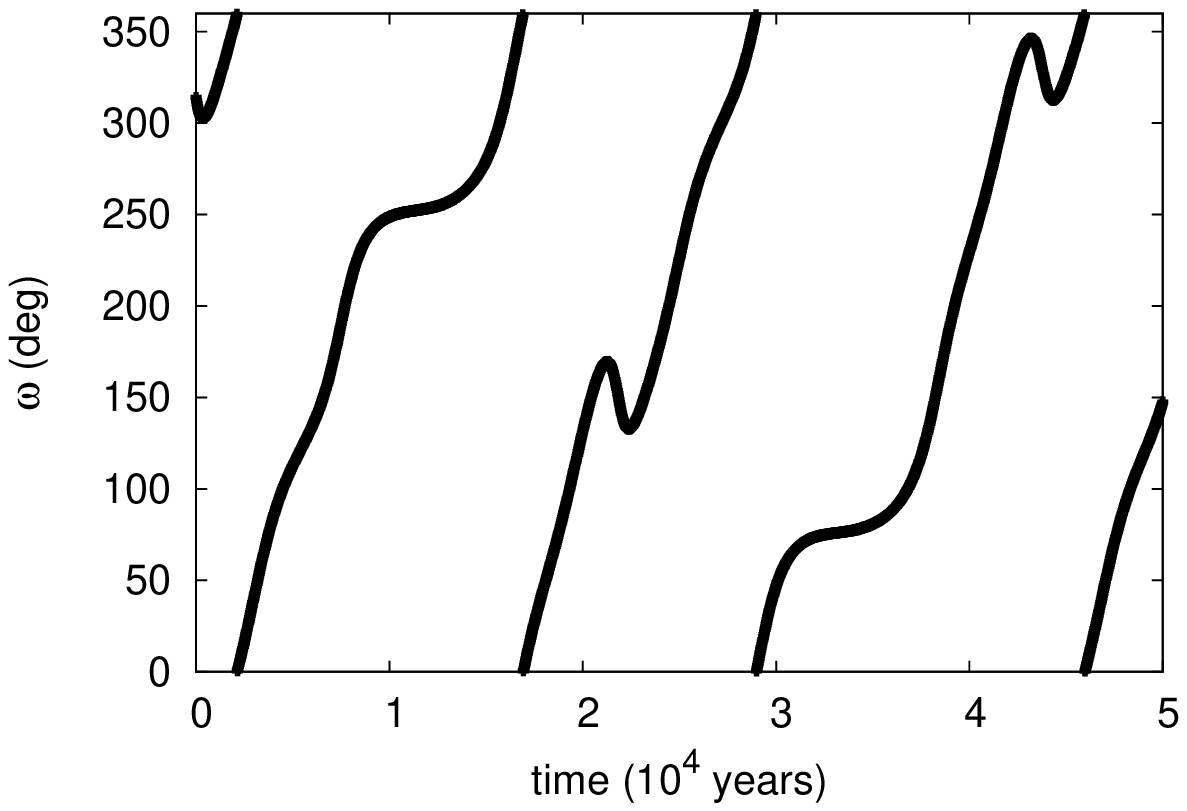}
\end{center}
\caption{\label{figura4}: 
Temporal evolution of the eccentricity, inclination, longitude of the ascending node and argument of pericentre
 of Romulus (left column) and Remus (right column).
 These are the results from the numerical integrations for the 4-body systems:
 Sylvia-Romulus-Sun-Jupiter (left) and  Sylvia-Remus-Sun-Jupiter (right) 
}
\end{figure*}

\section{The Evection Resonance and Collision}

The amplitude of oscillation of the satellites' inclinations can reach
higher values depending on the initial value of Jupiter's longitude of pericentre.
In Figure 6 we present the temporal evolution of the eccentricity and inclination 
of Romulus (left column) and Remus (right column),
considering different initial values of Jupiter's longitude of the pericentre.
 These are the results from the numerical integrations for the 5-body system  
Sylvia-Romulus-Remus-Sun-Jupiter, for
   $\varpi_{\rm Jup}=315^\circ$ (black),
$\varpi_{\rm Jup}=320^\circ$ (green), $\varpi_{\rm Jup}=10^\circ$ (red) and $\varpi_{\rm Jup}=15^\circ$ (blue).
We verified that for  $-40^\circ\leq\varpi_{\rm Jup}\leq 15^\circ$ the inclinations increase to higher values
and the eccentricities suddenly grow until the satellites collide with Sylvia

In those cases the satellites get caught in an evection resonance with Jupiter.
The evection resonance usually occurs when the period of the longitude of the pericentre of the satellite is very close to the orbital period of the perturber (Brouwer \& Clemence, 1961, Vieira Neto et al, 2006, Yokoyama et al. 2008). 
In the present study the resonant angle is given by the difference between the satellite's longitude of the pericentre and the mean longitude of Jupiter. 
In the first row of Figure 7 is presented an example of the evolutions of the evection angles of Romulus and Remus, for the case of initial $\varpi_{\rm Jup}=90^\circ$. 
One can note that when the evection angle starts to librate the corresponding eccentricity starts to grow until reach a collision with Sylvia.

\begin{figure*}
\begin{center}
\includegraphics[width=8.5cm]{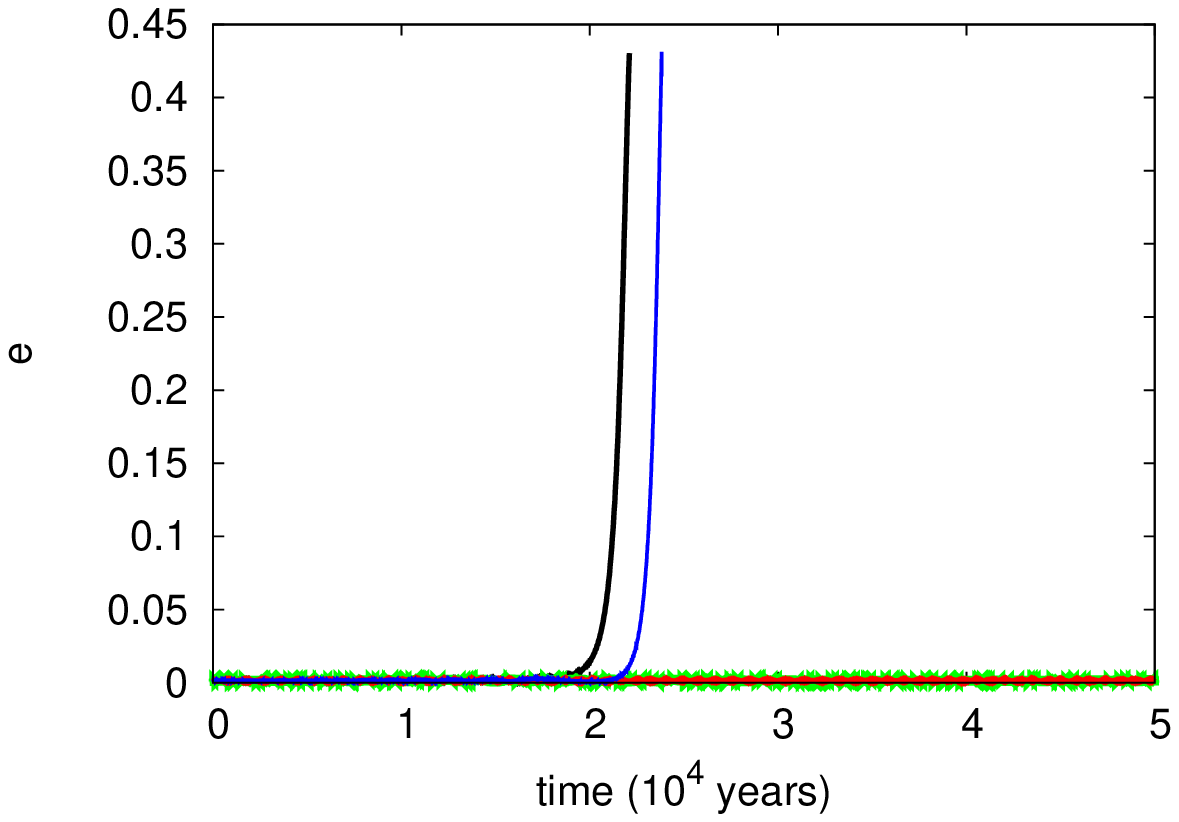}
\includegraphics[width=8.5cm]{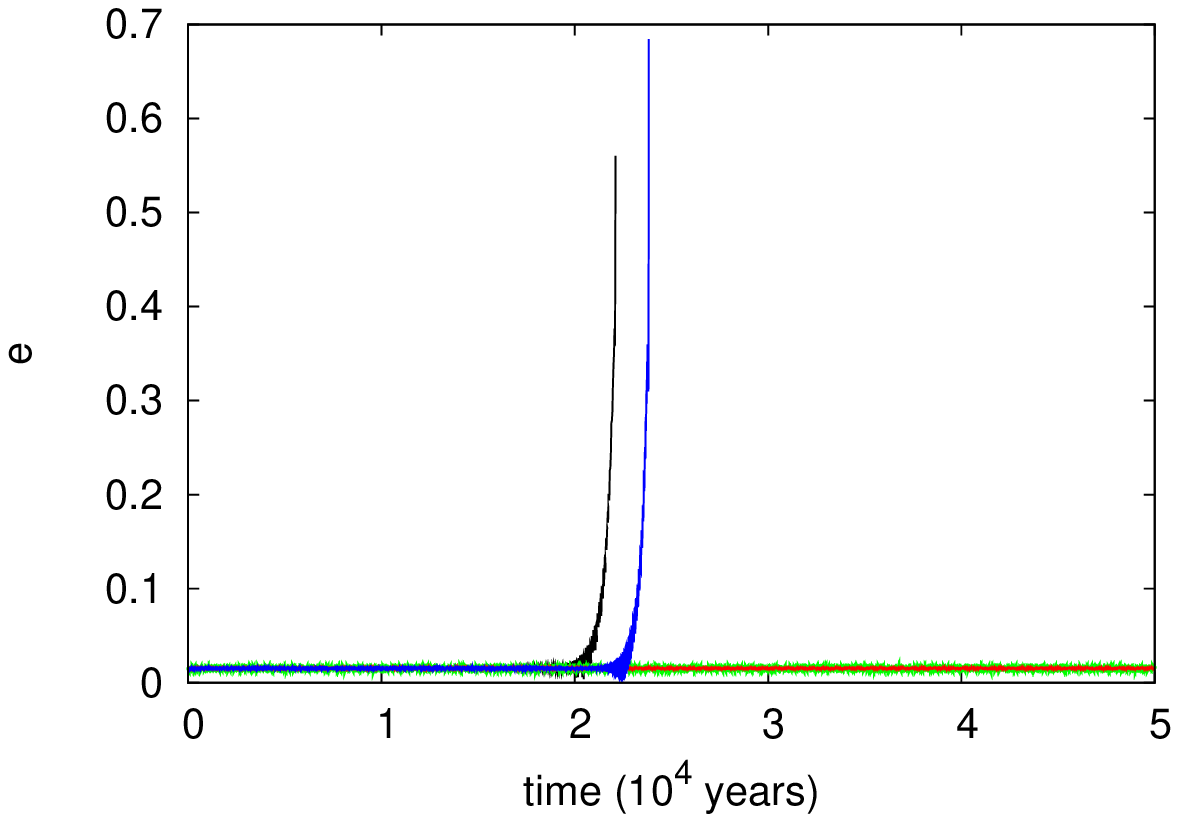}
\includegraphics[width=8.5cm]{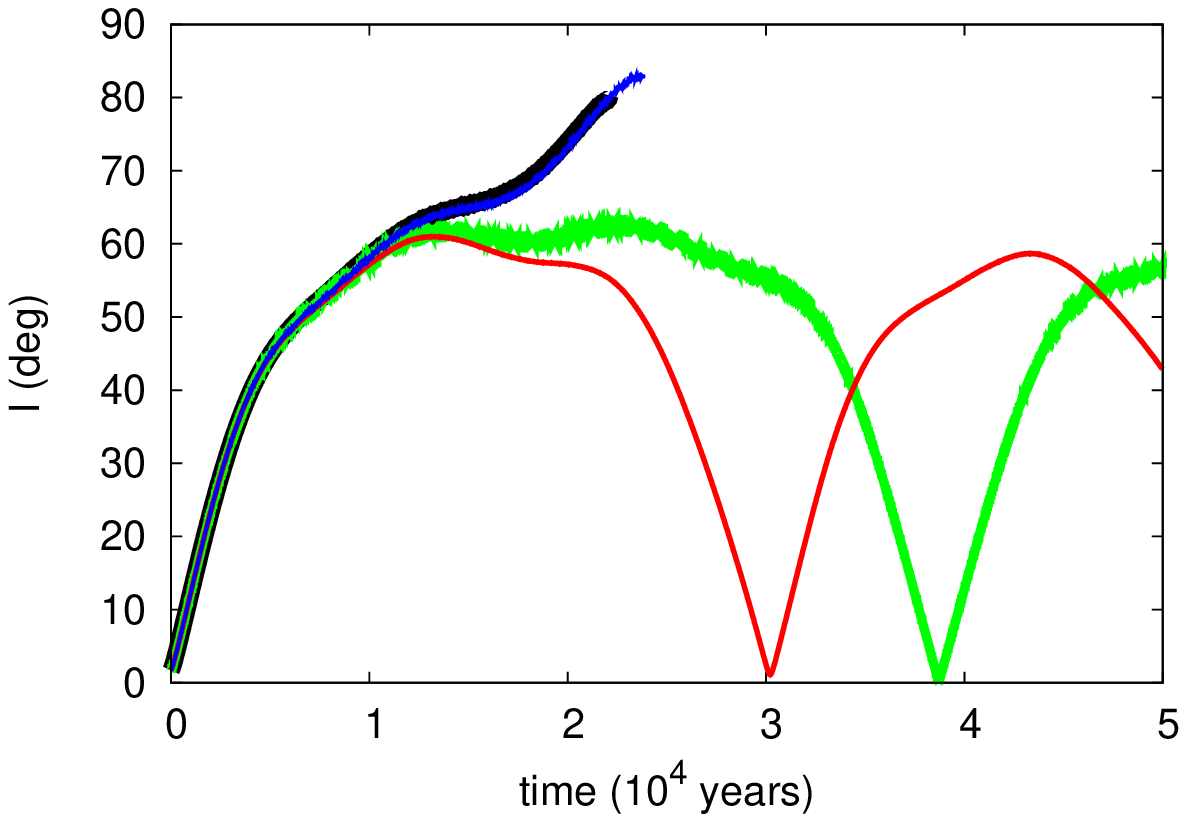}
\includegraphics[width=8.5cm]{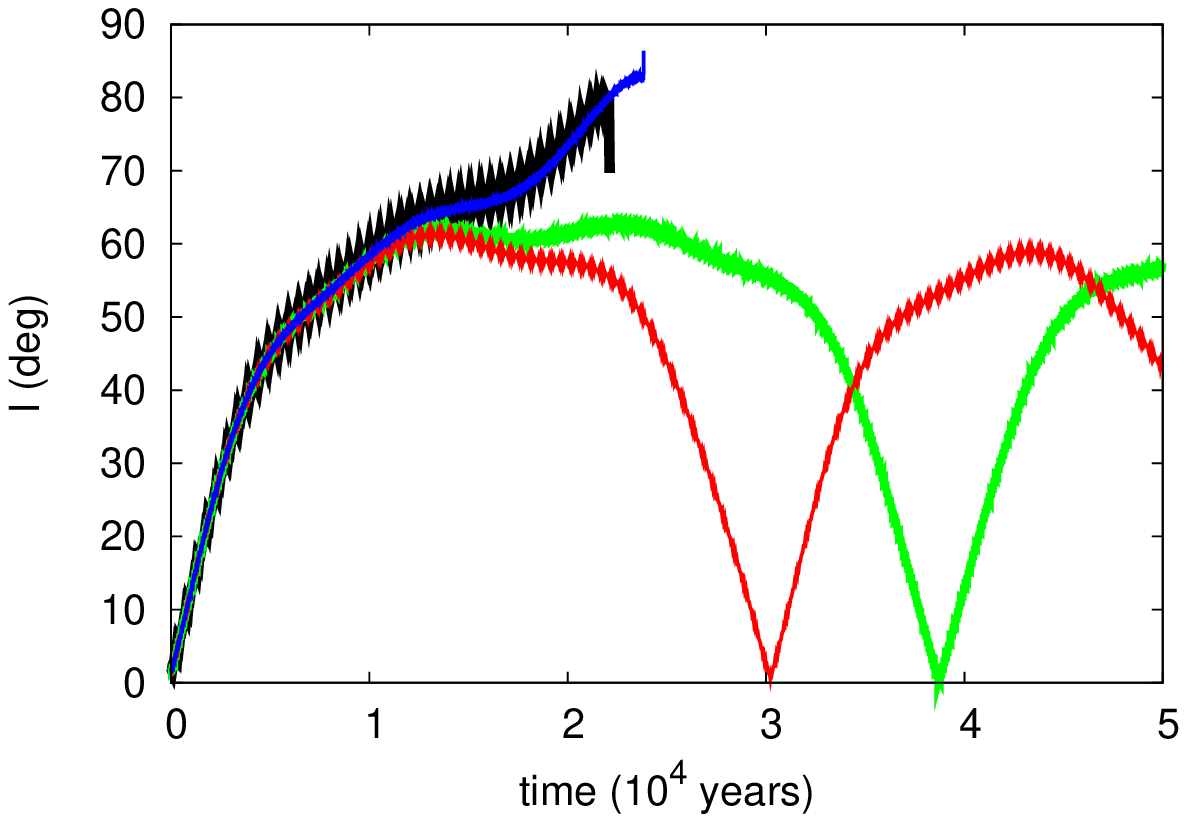}
\end{center}
\caption{\label{figura6}: 
Temporal evolution of the eccentricity and inclination of Romulus (left column) and Remus (right column).
 These are the results from the numerical integrations for the 5-body system  Sylvia-Romulus-Remus-Sun-Jupiter,
with different initial values of Jupiter's longitude of the pericentre: $\varpi_{\rm Jup}=315^\circ$ (black);
$\varpi_{\rm Jup}=320^\circ$ (green); $\varpi_{\rm Jup}=10^\circ$ (red); $\varpi_{\rm Jup}=15^\circ$ (blue).}
\end{figure*}

\begin{figure*}
\begin{center}
\includegraphics[height=6.0cm]{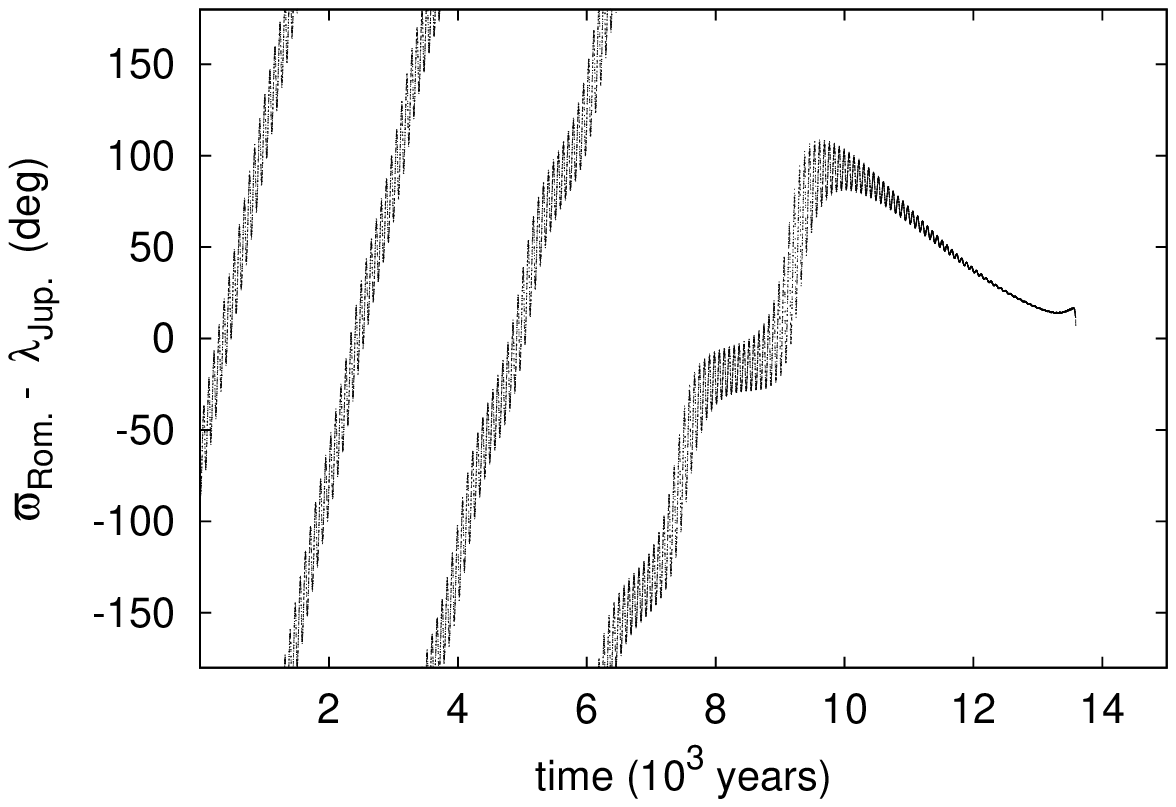}
\includegraphics[height=6.0cm]{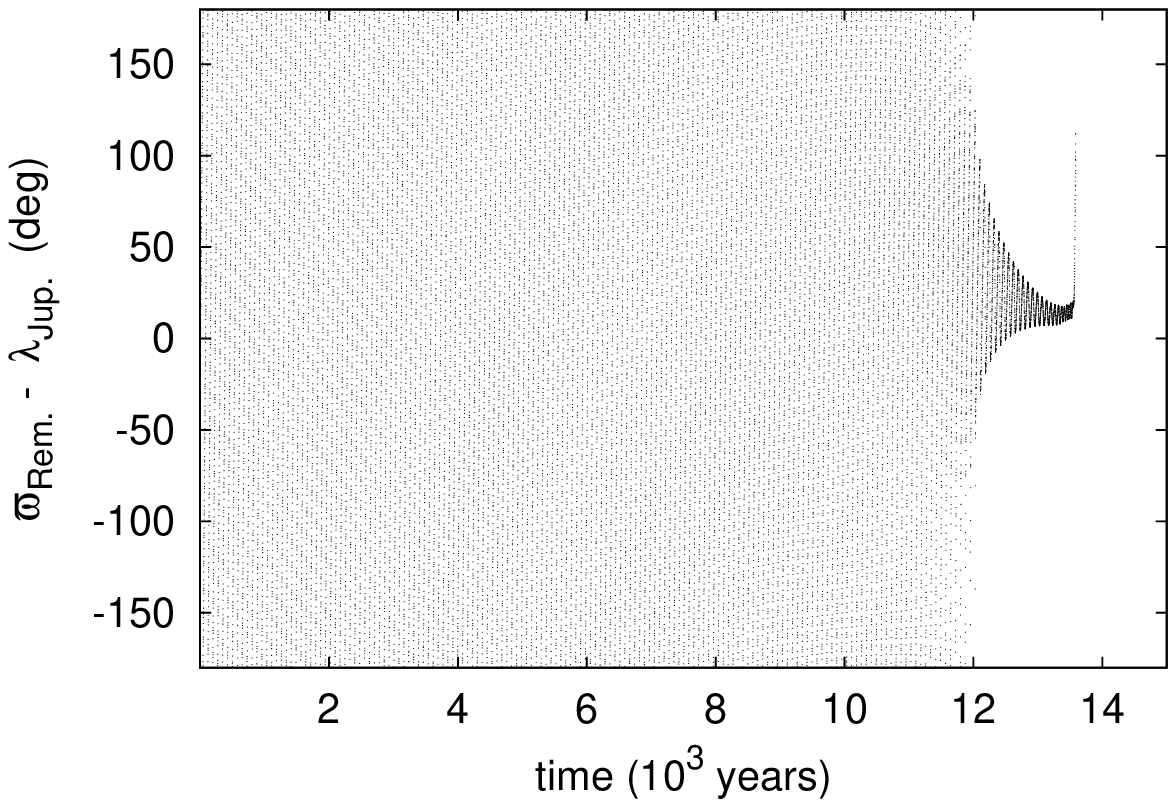}
\includegraphics[height=6.0cm]{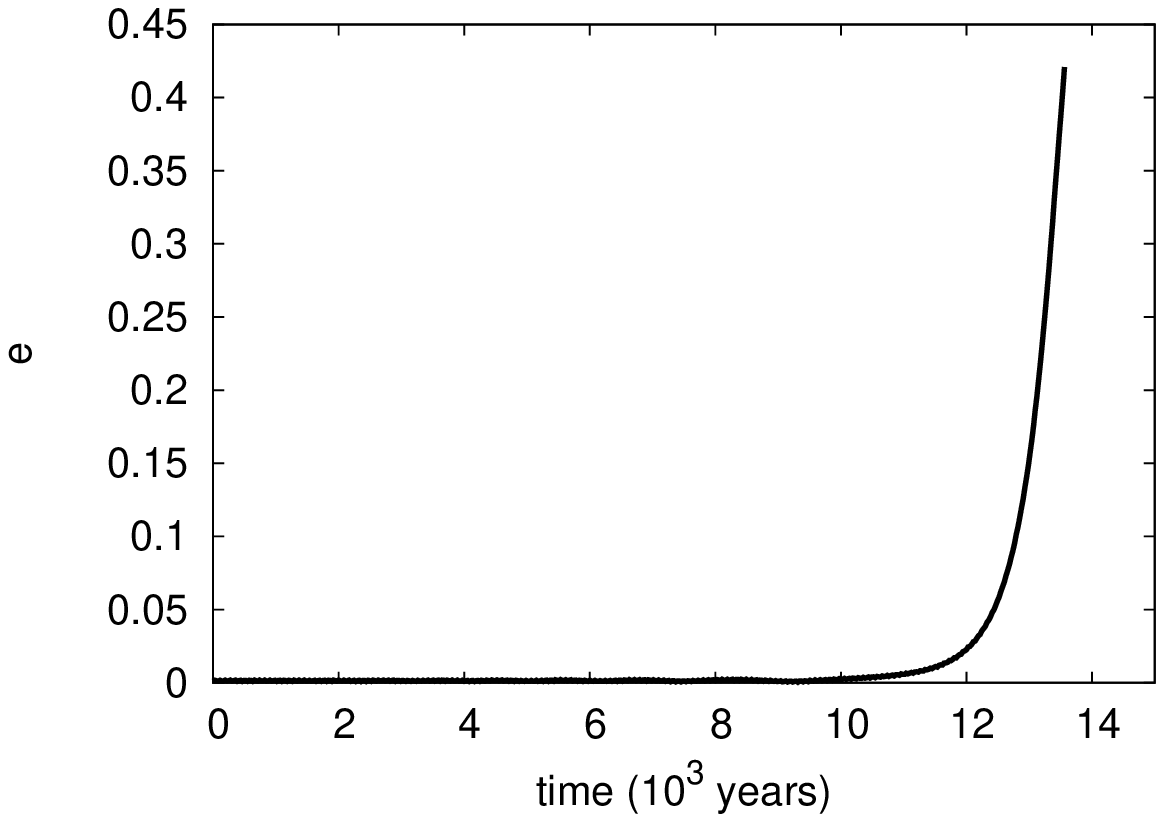}
\includegraphics[height=6.0cm]{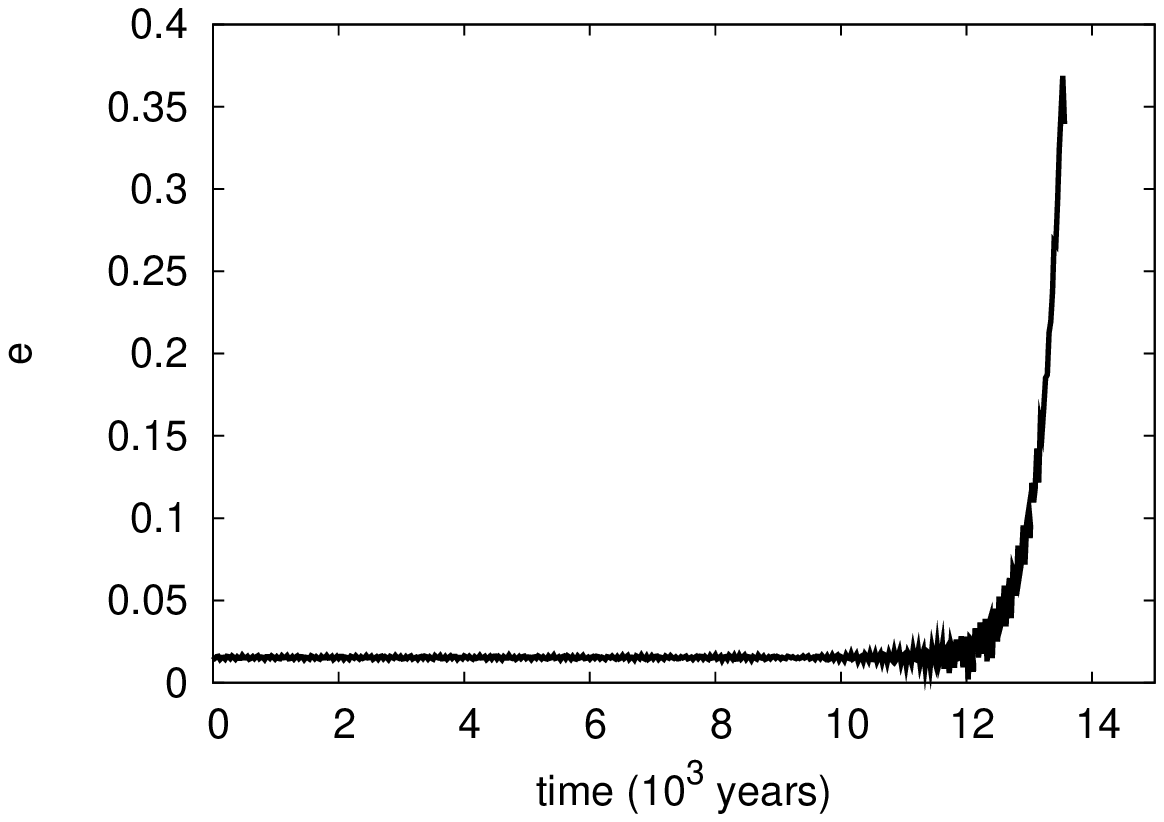}
\includegraphics[height=6.0cm]{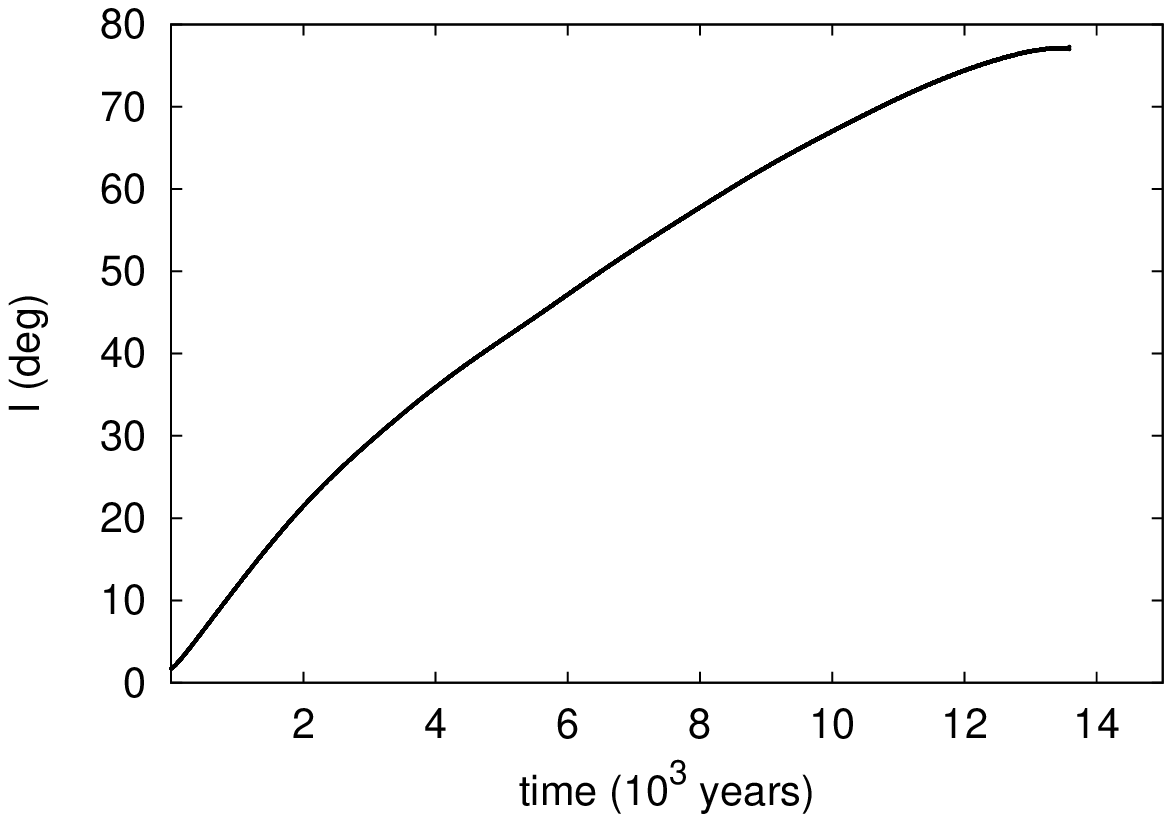}
\includegraphics[height=6.0cm]{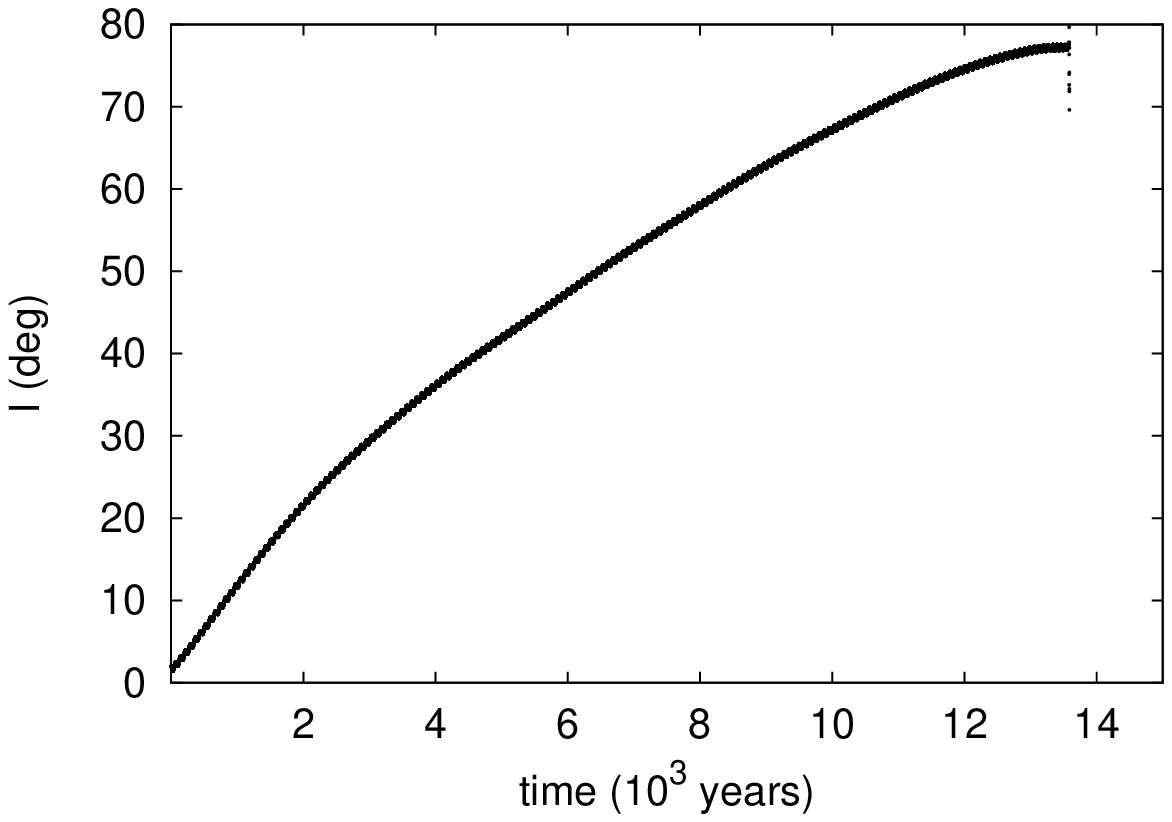}
\end{center}
\caption{\label{figura7}: Temporal evolution of the evection angle, eccentricity and inclination of Romulus (left column) and Remus (right column). These are the results from the numerical integrations for the 5-body system  Sylvia-Romulus-Remus-Sun-Jupiter, with initial value of the satellite's longitude of the pericentre equal to $90^\circ$.}
\end{figure*}

\section{Oblateness}

In this section the considered  subject is the oblateness of Sylvia.
First we discuss the limitations of the shape determination of Sylvia from the observational images.
Then, from secular perturbation theory (Ferraz-Mello, 1979), we analise the contribution of the Sun
in comparison to that of Sylvia's oblateness on the satellite's orbital inclination.
Finally, we present our numerical simulations for the 5-body system  Sylvia-Romulus-Remus-Sun-Jupiter including
the oblateness of Sylvia.

\subsection{Observational Constraints}

The $J_2$ of Sylvia was first measured from the analysis of both satellite orbits by Marchis et al. (2005). The knowledge for each of them of its mean motion and semi-major axis constrains simultaneously both the mass of Sylvia and its $J_2$ through the generalized Kepler's third law. The $J_2$ measured was thereby derived to be $0.175\pm 0.050$.  The global shape of Sylvia  was inferred from a convex inversion method of available ligthcurves by Kaasalainen et al. (2002). Furthermore, Marchis et al. (2006) recorded with the Keck Adaptive Optics system a resolved image of Sylvia's primary at an angular resolution of ~57 mas which  agreed very well with this 3D-shape model validating the convex model. Using the 3D-shape model and assuming an uniform mass distribution in the interior of the primary, we calculate a theoretical $J_2$ of 0.12, globally consistent with the measured value although slightly lower. The discrepancy may arise either from an interior heterogeneity or from the shape model itself which could have some local non-convexities, not taken into account in the inversion process, able to further increase the $J_2$ of Sylvia.

\subsection{Dynamical Effect}

In this section we consider the effect of Sylvia's oblateness ($J_2$) on the dynamics of Sylvia's satellites. We define $\Pi= I \exp{i \Omega}$ as a complex variable associated to the inclination ($I$) and longitude of the ascending node ($\Omega$). We use the indices 'o' for Romulus and 'e' for Remus and 'S' for the Sun. In what follows, we do not yet consider Sylvia's oblateness. Thus the secular equation for a satellite's $\Pi$ perturbed by the Sun is (Ferraz-Mello, 1979):

\begin{equation}
{d \Pi_o \over{d t} } = -i A_{S_o} \Pi_o + i A_{S_o} \Pi_S ,
\end{equation}

\noindent where

\begin{equation}
A_{S_o}={3 \over 4} {G m_S \over n_o a_s^3} ,
\end{equation}

\noindent with an equivalent equation for Remus. Here, $G$ is the gravitational constant, $m$ is a mass, $n$ a mean motion, $a$ is a semimajor axis and $i$ is the imaginary unit. The solution of Eq. (1), in the approximation $\Pi_S =$ constant,  is:

\begin{equation}
\Pi_o = K \exp{i g_o t} + \Pi_S ,
\end{equation}

\noindent where $K$ is a complex constant and $g_o = - A_{S_o}.$

We obtain $A_{S_o} \sim 0.1108\times 10^{-2} year^{-1}$, which corresponds to a period of $5673$ years, confirming the results of Section 3. For Remus, we find a period of $15012$ years by the same method. The forced component $\Pi_S$ induced by the Sun is thus responsible for the amplitude of the oscillation of $I$ around $20^{\circ}$, since $|\Pi_S| \sim 10^{\circ}$ and the initial conditions for $I_e$ and $I_o$ are near $0^{\circ}$.

We now add the second satellite to the secular equations. To do this, we consider a complex vectorial variable $\Pi = (\Pi_o,\Pi_e)$. The equation for the secular variation of $\Pi$ (Ferraz-Mello,1979) is similar to Eq. (1):

\begin{equation}
{d \Pi_o \over{d t} } = i\; A \;\Pi + i\; B\; \Pi_S
\end{equation}
where $A$ is a $2 \times 2$ matrix with elements:

\begin{eqnarray}
&&A(1,1)=-(A_e+A_{S_o}), \\
&&A(1,2)=A_e ,\\
&&A(2,1)=A_o, \\
&&A(2,2)=-(A_o+A_{S_e},) 
\end{eqnarray}
and $B$ is a $2 \times 1$ vector with components:

\begin{eqnarray}
&&B(1)=A_{S_o},  \\
&&B(2)=A_{S_e}, 
\end{eqnarray}
where

\begin{eqnarray}
&&A_e = {1 \over 4} {G m_e L \over n_o a_o^2},  \\
&&A_o = {1 \over 4} {G m_o L \over n_e a_e^2},  \\
&&L = {a_e \over a_o^2} b_{3/2}^{(1)}, 
\end{eqnarray}
where $b_{3/2}^{(1)}$ is a Laplace coefficient (Ferraz-Mello, 1979). The solution of Eq. (4) is quite similar to the one-dimensional case:

\begin{equation}
\Pi = N.E + C.\Pi_S,
\end{equation}
where $N$ is a $2 \times 2$ complex matrix whose rows are eigenvectors of matrix $A$ and the corresponding eigenvalues are $g_1$ and $g_2$ in $E=(\exp{i g_1},\exp{i g_2})$. $C$ is a 'unitary' vector $(1,1)$. This results from the fact that  $A(j,1)+A(j,2)+B(j)=0$, for $j=1,2$. As far as the forced component is concerned, both Romulus and Remus  share a common one, given by the inclination of the Sylvia-Sun orbital plane with respect to Sylvia's equator. Thus the $20^{\circ}$ amplitude of oscillation of $I_e$ and $I_o$ are the same as in the case of just one satellite (Fig. 3). As to the frequencies of oscillation, they are the eigenfrequencies of matrix $A$ and we find them to be:

\begin{eqnarray}
&&g_1=-0.108 \times 10^{-2} year^{-1},  \\
&&g_2=-0.832 \times 10^{-1} year^{-1}. 
\end{eqnarray}
These correspond to periods of around $75$ and $5820$ years, a long period forced by the Sun and a short period due to the satellites themselves. This short period is similar to the secular one defined by just the two satellites around Sylvia and the long period (similar to the period of Romulus in the Sylvia-Romulus-Sun problem) is shared by both satellites as shown in Fig. (3) (third and fourth row).

We now introduce the oblateness of Sylvia defined by its $J_2$. To better understand its effect we first include it in the secular equations for just one satellite and the Sun. In this  way, Eq. (1) becomes:

\begin{equation}
{d \Pi_o \over{d t} } = -i (A_{S_o}+A_{J_o}) + i A_{S_o} \Pi_S
\end{equation}
where

\begin{equation}
A_{J_o} = {3 \over 2} {G J_2 m_{sy} b^2 \over n_o a_o^5}.
\end{equation}
Here, $m_{sy}$ is Sylvia's mass and $b$ is Sylvia's equatorial radius. Now the solution of (17) is:

\begin{equation}
\Pi_o = K \exp{i g_o t} + c \Pi_S,
\end{equation}
where $K$ is a constant and:

\begin{eqnarray}
&&g_o = -(A_{S_o}+A_{J_o}),  \\
&&c = {A_{S_o} \over A_{S_o} + A_{J_o}}. 
\end{eqnarray}

The value of $A_{J_o}$, with $J_2 = 0.17$ is $1.7401 year^{-1}$. Since $A_{J_o} \gg A_{S_o}$, first $I_o$ will oscillate with a much higher frequency and second, the forced component, now multiplied by a very small constant $c$, will almost vanish.

The equations for both satellites, the Sun and including Sylvia's oblateness will be like Eq (4), but the diagonal elements of matrix $A$ are now:

\begin{eqnarray}
&&A(1,1)=-(A_e+A_{S_o}+A_{J_o}), \\
&&A(2,2)=-(A_o+A_{S_e}+A_{J_e}).
\end{eqnarray}

This will have important consequences on the eigenvalues (proper frequencies) and the forced components. They now become:

\begin{eqnarray}
&&g_1=-1.745\; year^{-1},\\
&&g_2=-17.04\; year^{-1},  \\
&& C = (6.35 \times 10^{-4}, 2.75 \times 10^{-5}),  
\end{eqnarray}
where $C = A^{-1} B$. Thus the consequence of Sylvia's oblateness is to critically stabilize the satellite's orbits, since it introduces a frequency that is much faster than those generated by the other gravitational perturbations.


\subsubsection{The effect of Jupiter}

The direct effect of Jupiter on Sylvia's satellites is negligible as compared with the Sun's effect. However, by considering Jupiter, Sylvia's orbital plane is no longer fixed but now precess with a frequency given by:

\begin{equation}
\mu = {1 \over 4} {G\; m_J\; L_J \over n_{Sy}\; a_{Sy}^2},
\end{equation}
where the subscript 'J' refers to Jupiter and 'Sy' to Sylvia and 
\begin{equation}
L_J = {a_{Sy} \over a_J} b_{3/2}^{(1)}
\end{equation}

Since Jupiter has a nonzero inclination with respect to the reference plane (ecliptic) Sylvia's inclination-node complex variable $\Pi$ will be the sum of a constant forced component and a proper component precessing with frequency $\mu$. Thus:
\begin{equation}
\Pi_{Sy} = \Pi_J + K \exp{-i \mu t}
\end{equation}
where $K$ is a constant to be determined as far as one knows $\Pi_{Sy}$ for a specific time. 

Now to solve equations (1), (4) and (17), we must replace the constant $\Pi_S$ by $\Pi_{Sy}$ given by equation (29). The solution for each equation and a specific satellite can be generally given by:
\begin{equation}
\Pi_o = \Pi_{o_p} + k_1 \; K exp{-i \mu t} + k_2 \Pi_J,
\end{equation}
where $\Pi_{o_p}$ is the proper component which will not change from the case without Jupiter, and
\begin{eqnarray}
&&k_1 = {A_{S_o} \over A_{S_o}+A_J-\mu} \; K ,\\
&&k_2 = {A_{S_o} \over A_{S_o}+A_J},  
\end{eqnarray}
for the case of just one satellite here represented by Romulus (index 'o'). When both satellites are included then $k_1$ and $k_2$ are two dimension vectors given by:
\begin{eqnarray}
&&k_1 = -(A + \mu D)^{-1} B ,\\
&&k_2 = -A^{-1} B,  
\end{eqnarray}
where $A$ and $B$ are matrices already defined for the case without Jupiter and 'D' is the identity matrix.

This powerfull effect of the oblateness was also explored in our numerical simulations.
In Figure 8 we present the results from the numerical integrations for the 5-body system  Sylvia-Romulus-Remus-Sun-Jupiter, with different values for $J_2$ of Sylvia:  i) In purple, $J_2=0$. ii) In blue,  $J_2=10^{-3}\,J_{2 {\rm Sylvia}}$. iii) In green,  $J_2=10^{-2}\,J_{2 {\rm Sylvia}}$. iv) In red,  $J_2= J_{2 {\rm Sylvia}}$.
The temporal evolution of the inclination and the longitude of the ascending node of Romulus (left column) and Remus (right column) clearly show that the value of $J_2$ has to be much smaller (less than one thousand times) than the currently estimated value for Sylvia, in order to display any effect from other bodies' perturbation.

\begin{figure*}
\begin{center}
\includegraphics[width=8.5cm]{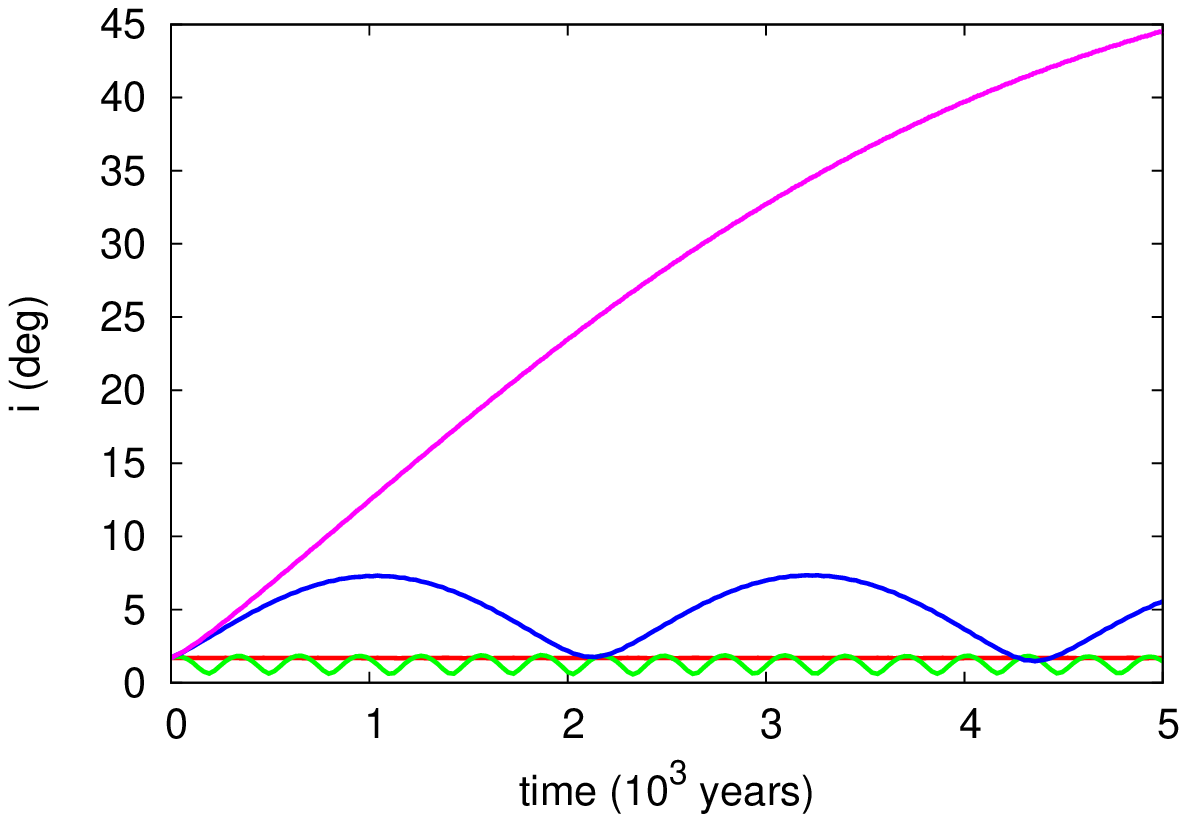}
\includegraphics[width=8.5cm]{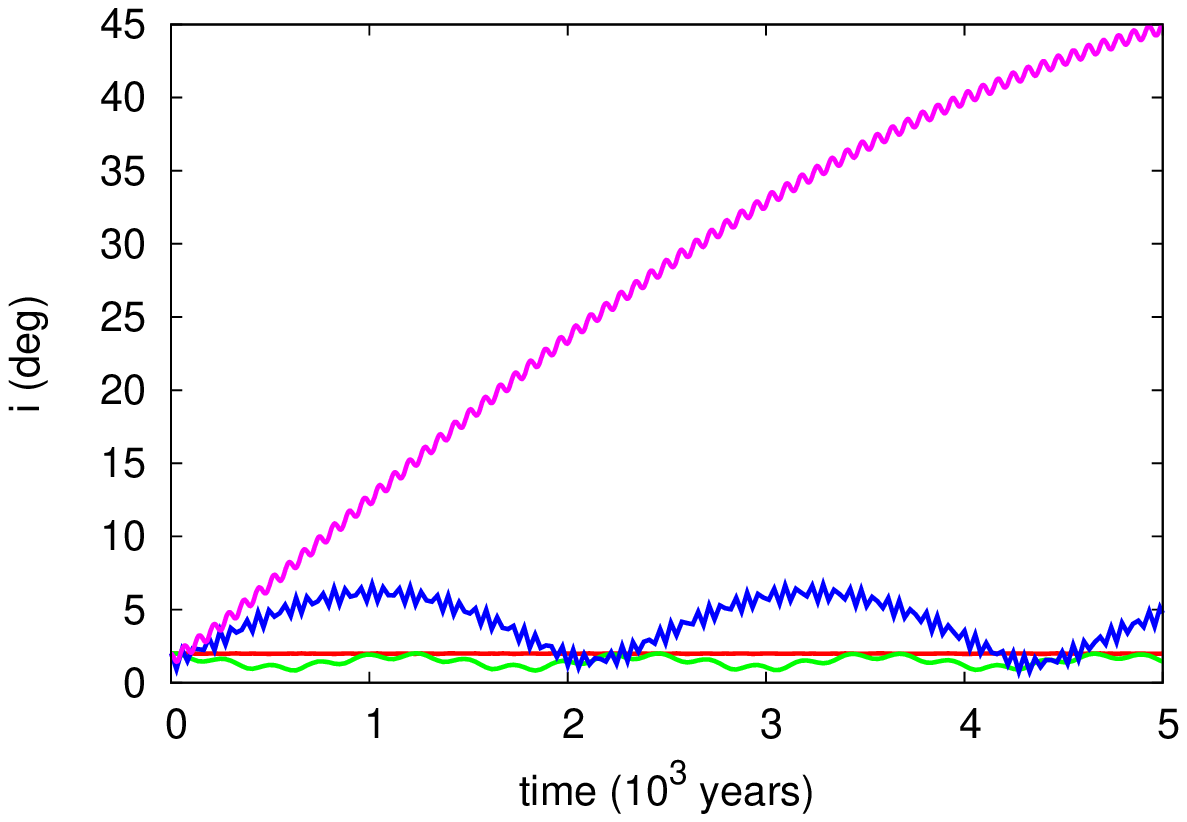}
\includegraphics[width=8.5cm]{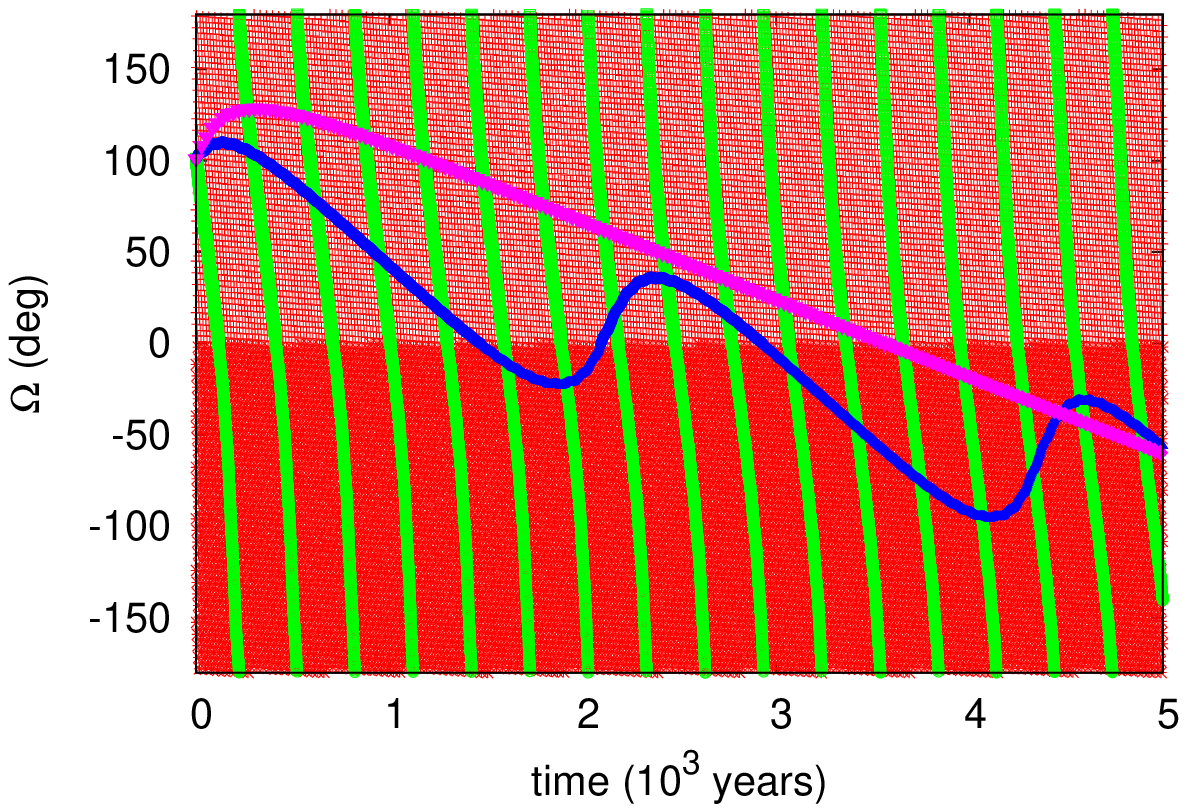}
\includegraphics[width=8.5cm]{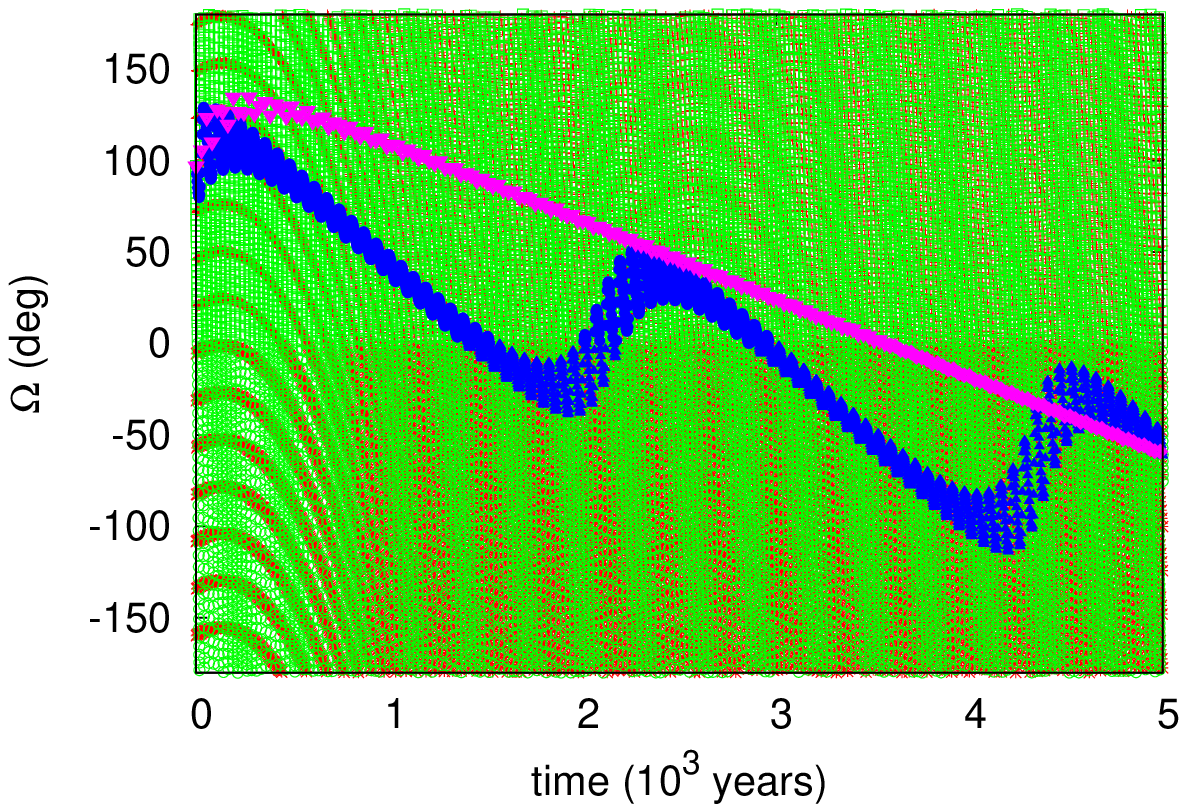}
\end{center}
\caption{\label{figura8}: 
Temporal evolution of the inclination and the longitude of the ascending node of Romulus (left column) and Remus (right column).
These are the results from the numerical integrations for the 5-body system  Sylvia-Romulus-Remus-Sun-Jupiter,
with different values for $J_2$ of Sylvia.  i) In purple, $J_2=0$. ii) In blue,  $J_2=10^{-3}\,J_{2 {\rm Sylvia}}$. iii) In green,  $J_2=10^{-2}\,J_{2 {\rm Sylvia}}$. iv) In red,  $J_2= J_{2 {\rm Sylvia}}$.
}
\end{figure*}

\section{Conclusions}

We have explored the dynamics of the satellites of asteroid (87) Sylvia.
When the oblateness of Sylvia is not taken into account, the satellites
present a very interesting dynamical evolution.
The perturbations from the Sun and Jupiter introduce a huge increase on
the satellites orbital inclinations.
Depending on the initial longitude of the pericentre of Jupiter, the satellites
can be captured in a kind of evection resonance. The longitudes of pericentre of the satellites
get locked to the mean longitude of Jupiter.
Such resonance forces the satellites eccentricities to grow exponentially and they eventually
collide with Sylvia.
It is also noted that the evolutions of Romulus and Remus are very similar.
This is due to a secular resonance between them, which is caused by the perturbations from Sun and Jupiter.
Finally, the complete stability of the satellites is guaranteed by the oblateness of Sylvia.
We show that even just one thousand of the current value of $J_2$ would be enough to keep
the satellites in stable orbits.

There are other triple asteroid systems in the main belt with similar characteristics, (45) Eugenia (Marchis et al 2007) and (216) Kleopatra (Marchis et al 2008), which will be studied adopting a similar approach.

\section{Acknowledgments}

This work was supported by the Brazilian agencies FAPESP, CNPq and CAPES.
F.M. is supported by National Aeronautics and Space Administration issue through the Science Mission
Directorate Research and Analysis programs number NNX07AP70G.

\section{References}

Brouwer, D. \& Clemence, G.M. 1961, Methods of Celestial Mechanics, Academic Press, London.

Everhart, E. 1985, in Dynamics of Comets: Their Origin and
Evolution, ed. A. Carusi, \& G. Valsecchi (Dordrecht: Reidel), 185.

Ferraz-Mello, S. 1979, Dynamics of the Galilean Satellites: An Introductory Treatise, 
Instituto Astron\^ omico e Geof\' \i sico, Universidade de S\~ ao Paulo, S\~ ao Paulo.

Kaasalainen, M., Torppa, J., Piironen, J. 2002, Icarus, 159, 369, 2002

Marchis, F., Descamps, P., Berthier, J., Emery, J. P. 2008, IAU Circ. no. 8980.

Marchis, F., Baek, M., Descamps, P., Berthier, J., Hestroffer, D., Vachier, F. 2007, IAU Circ. no. 8817.

Marchis, F., Kaasalainen, M., Hom, E.F.Y., Berthier, J., Enriquez, J., Hestroffer, D., Le Mignant D., de Pater, I. 2006, Icarus, 185, 39.

Marchis, F., Descamps, P., Hestroffer, D., Berthier, J., Brown, M. E., Margot, J.-L. 2005, IAU Circ. no. 8582.

Marchis, F., Descamps, P., Hestroffer, D., Berthier, J. 2005, Nature, 436, 822.

Murray, C.D. \& Dermott, S.F. 1999, Solar System Dynamics, Cambridge University Press, Cambridge.

Vieira Neto, E., Winter, O.C., Yokoyama, T. 2006, A\&A, 452, 1091.

Yokoyama, T, Vieira Neto, E., Winter, O.C., Sanches, D.M., Brasil, P.I. 2008, Mathematical Problems in Engineering, 2008, ID 251978 (doi:10.1155/2008/251978).

\end{document}